# Writing and detecting topological charges in exfoliated $Fe_{5-x}GeTe_2$

*Alex Moon[§,£], Yue Li[¤], Conor McKeever[¥], Brian W. Casas[§], Moises Bravo[ɩ], Wenkai Zheng[§,£], Juan Macy[§,£], Amanda K. Petford-Long[¤,ƺ], Gregory T. McCandless[ɩ], Julia Y. Chan[ɩ], Charudatta Phatak[¤,ƺ], Elton J. G. Santos[¥,ǂ,*], and Luis Balicas[§,£,*]*

*[§]National High Magnetic Field Laboratory, 1800 E. Paul Dirac Dr. Tallahassee, FL 32310, USA.*

*[£]Department of Physics, Florida State University, 77 Chieftan Way, Tallahassee, FL 32306, USA.*

*[¤]Materials Science Division, Argonne National Laboratory, Lemont, 60439, IL, USA.*

*[¥]Institute for Condensed Matter and Complex Systems, School of Physics and Astronomy, The University of Edinburgh, EH9 3FD, UK.*

*[ɩ]Department of Chemistry and Biochemistry, Baylor University, Waco, TX 76798, USA.*

*[ƺ]Department of Materials Science and Engineering, Northwestern University, Evanston, 60208, IL, USA.*

*[ǂ]Higgs Centre for Theoretical Physics, The University of Edinburgh, EH9 3FD, UK.*

ABSTRACT: $Fe_{5-x}GeTe_2$ is a promising two-dimensional (2D) van der Waals (vdW) magnet for practical applications, given its magnetic properties. These include Curie temperatures above room temperature, and topological spin textures – TST (both merons and skyrmions), responsible for a pronounced anomalous Hall effect (AHE) and its topological counterpart (THE), which can be harvested for spintronics. Here, we show that both the AHE and THE can be amplified considerably by just adjusting the thickness of exfoliated $Fe_{5-x}GeTe_2$, with THE becoming observable even in zero magnetic field due to a field-induced unbalance in topological charges. Using a complementary suite of techniques, including electronic transport, Lorentz transmission electron microscopy, and micromagnetic simulations, we reveal the emergence of substantial coercive fields upon exfoliation, which are absent in the bulk, implying thickness-dependent magnetic interactions that affect the TST. We detected a 'magic' thickness *t* ~30 nm where the formation of TST is maximized, inducing large magnitudes for the topological charge density (~6.45 x $10^{20}$ $cm^{-2}$), and the concomitant anomalous ($\rho_{xy}^{\mathrm{A,max}} \cong$ 22.6 μΩ cm) and topological ($\rho_{xy}^{\mathrm{u,T}} \cong$ 15 μΩ cm) Hall resistivities at $T$ ~ 120 K. These values for $\rho_{xy}^{\mathrm{A,max}}$ and $\rho_{xy}^{\mathrm{u,T}}$ are higher than those found in magnetic topological insulators and, so far, the largest reported for 2D magnets. The hitherto unobserved THE under zero magnetic field could provide a platform for the writing and electrical detection of TST aiming at energy-efficient devices based on vdW ferromagnets.

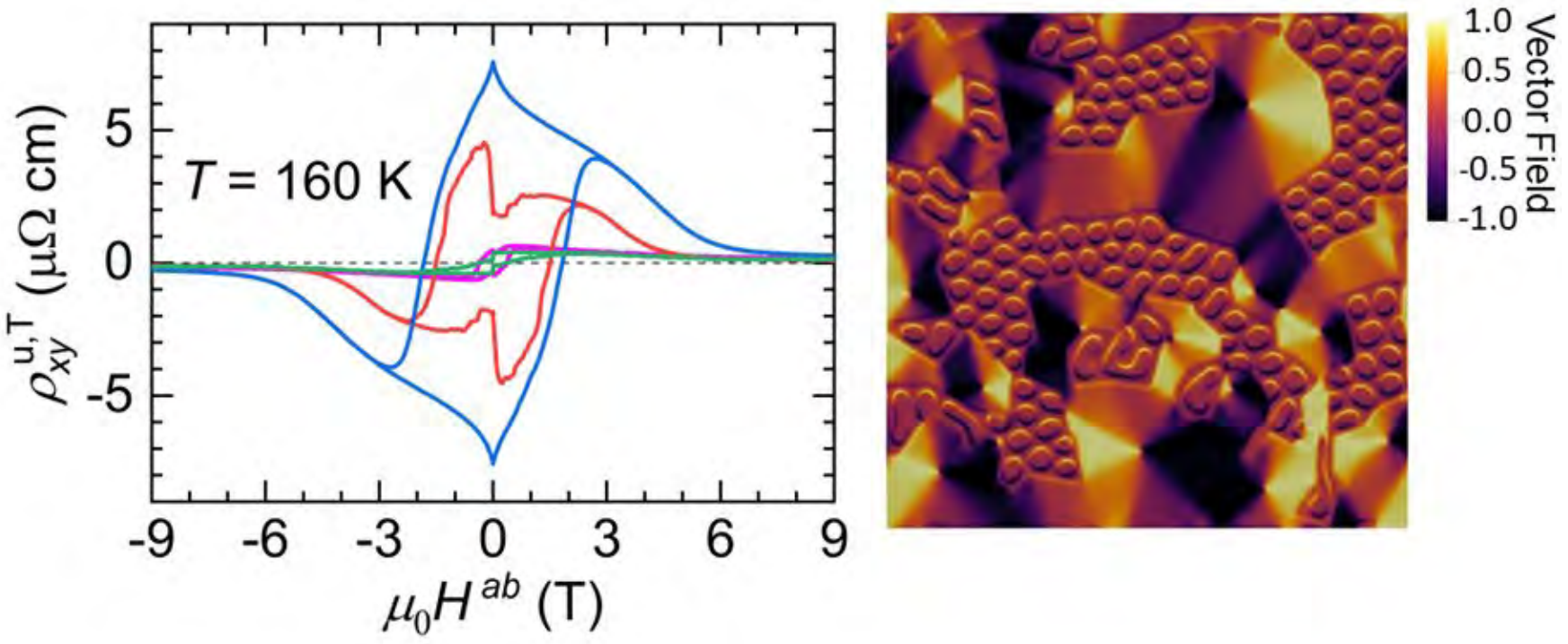

The study of magnetism in the two-dimensional (2D) limit has been pivotal for the development of critical phenomena and strongly correlated phases in ultrathin compounds[1-3]. Such van der Waals (vdW) magnetism has seen a recent resurgence due to the advent of layered magnetic compounds displaying a magnetic ground state even when exfoliated down to the monolayer limit[4-7]. Since 2D magnetism is not ruled out by the non-existence of crystalline magnetic anisotropy, as recently demonstrated[8], several materials have been explored towards for fundamental interest and practical applications, in particular semiconducting tri-halides[9]. Among known vdW layered magnets, the metallic ferromagnets belonging to the $Fe_{n-x}GeTe_2$ family, and its doped variants, display the highest known Curie temperature ($T_c$) with $T_c$ ~ (205 K – 230 K)[8] for $n$ = 3, ~278 K[10] for $n$ = 4, (~270 K – 330 K)[11] for $n$ = 5, and > 400 K[12] for Ni-doped $Fe_5GeTe_2$. In these compounds, not only is magnetism observed in very thin layers, but there is also tunability *via* a perpendicular electric field. In tri-layered $Fe_{3-x}GeTe_2$, $T_c$ increases from ~ 100 K up to room temperature *via* the use of ionic liquid gating[13]. Meanwhile, the ground state of two unit cell thick $Fe_{5-x}GeTe_2$[14] and bulk $Fe_{3-x}GeTe_2$[15] switches from ferromagnetic to antiferromagnetic upon solid proton electrolyte gating. Despite being centrosymmetric, implying *a priori* the absence of the Dzyaloshinskii–Moriya interaction[16] (DMI), these compounds are prone to display topological spin textures, such as skyrmions[17-19], merons[19-21], and complex domain boundaries between stripe domains[22].

Topological spin textures such as skyrmions, are vortex-like nanometric spin textures that carry an integer topological number, or topological charge, describing how many times the magnetic moments composing it wrap around a sphere. In contrast, merons are characterized by a half integer topological charge. Such textures are characterized by a finite value of the scalar field spin chirality, $\chi_{ijk} = \vec{S_i} \cdot (\vec{S_j} \times \vec{S_k})$, which is a fictitious magnetic field that bends the electronic orbits, generating the so-called topological Hall effect (THE)[23-26]. In magnetic compounds, the

THE is usually treated as an additive contribution to the conventional and anomalous Hall responses:

$$\rho_{xy} = \rho_{xy}^{\mathrm{N}} + \rho_{xy}^{\mathrm{A}} + \rho_{xy}^{\mathrm{T}} \qquad (1)$$

where $\rho_{xy}^{\mathrm{N}} = R_0 B$ is the conventional Hall response, with $R_0$ being the Hall coefficient, $\rho_{xy}^{\mathrm{A}}$=$S_H M \rho_{xx}^{n}$ the anomalous Hall effect (AHE) term, with $S_H$ being the anomalous Hall constant, $B$ the induction field, $M$ the magnetization, $\rho_{xx}^{n}$ a power of the magnetoresistivity, and $\rho_{xy}^{\mathrm{T}}$ the THE contribution. Through a semiclassical theoretical approach that includes the solution of the Boltzmann equation, it was recently shown that these contributions are indeed additive, with the intrinsic anomalous Hall effect resulting from the Berry curvature (acquired by charge carriers as they move) in momentum-space, and the topological term from the Berry curvature in real-space[27]. Based on a scaling analysis proposed by Ref. [28], it was recently argued that the anomalous Hall effect observed in $Fe_{5-x}GeTe_2$ is dominated by the intrinsic contribution[29]. This implies that $\rho_{xy}^{\mathrm{A}}$ should be dominated by either the Berry phase resulting from carriers scattering from the topological spin textures, or the Karplus Luttinger term[30], instead of extrinsic mechanisms such as skew scattering or side jumps[31].

Given that spin chirality is a scalar and not a vector field, it can deflect moving charges regardless of the orientation of an applied external magnetic field. This point is illustrated by the observation of an antisymmetric Hall-like signal in both $Fe_{3-x}GeTe_2$ (Ref.[32]) and $Fe_{5-x}GeTe_2$ (Ref.[19]) when the external magnetic field is oriented along the electrical currents, or in the absence of Lorentz force. In both systems, this unconventional THE-like signal is found to be magnetic-field dependent, given that the field alters the spin textures (and associated spin chirality) responsible for this unconventional THE or $\rho_{xy}^{\mathrm{u,T}}$. In the case of $Fe_{3-x}GeTe_2$, $\rho_{xy}^{\mathrm{u,T}}$ peaks at low temperatures in the vicinity of $\mu_0 H \cong 4.5$ T, with this behavior being mimicked by both the unconventional topological Nernst response, $S_{xy}^{\mathrm{u,T}}$, and the topological thermal Hall response, $\kappa_{xy}^{\mathrm{u,T}}$, measured with the thermal gradient $\nabla T$ aligned along $\mu_0 H$ in the $ab$-plane[32]. Here, the term "unconventional" refers to the unconventional measurement configurations.

An intriguing aspect of $Fe_{5-x}GeTe_2$, and its doped variants, is the fact that its magnetic domains, and associated spin textures, can be modified *via* simple exfoliation[21, 29, 33, 34]. Ref. [21] attributes such magnetic transformations to thickness-dependent changes in the relative strengths of the exchange constants, magnetic uniaxial anisotropy, and magnetic dipole interaction. This leads to thickness-dependent magnetic phase diagrams[21], and, apparently, also to the observation of skyrmions in $(Fe_{0.5}Co_{0.5})_{5-x}GeTe_2$ in a narrow range of thicknesses[34], *i.e.*, from ~ 0.1 μm to ~ 1 μm. Seemingly, exfoliation increases the relevance of the dipolar interactions in detriment of the DMI, increasing also the magnetic fluctuations that lead to lower Curie temperatures, effectively making the interplanar exchange coupling less relevant to 2D magnets. This modulates the spin texture, for instance, in $(Fe_{0.5}Co_{0.5})_{5-x}GeTe_2$, the Néel skyrmion size *d* was reported to behave as $d \propto t^{1/2}$, with *t* being the thickness, hence following Kittel's law[34]. It also leads to the emergence of hysteresis in both the magnetoresistivity and AHE in exfoliated samples[29] although this is absent in bulk crystals (see, Fig. S1). In thin lamellas of $Fe_{1.9}Ni_{0.9}Pd_{0.2}P$, square shaped antiskyrmions are observed to transition to elliptical like skyrmions as the lamella thickness *t* is reduced to $t$ ~ 50 nm and claimed to result from the increased relevance of the dipolar interaction, relative to the DMI, at *t* is reduced[35]. It is, therefore, pertinent to ask if, and how, the spin textures evolve as a function of layer thickness and if such evolution might affect the unconventional THE[20, 36] observed in $Fe_{5-x}GeTe_2$.

Here, we evaluate the unconventional THE ($\rho_{xy}^{\mathrm{u,T}}$) and AHE ($\rho_{xy}^{\mathrm{A}}$) responses of $Fe_{5-x}GeTe_2$ as a function of both the crystal thickness *t* within 12 nm ≤ *t* ≤ 65 nm and temperature. For fields parallel to electrical currents flowing along a planar direction, we find a coercive field $\mu_0 H_c^{ab} > 10$ $\mu_0 H_c^c$, where $\mu_0 H_c^{ab}$ and $\mu_0 H_c^c$ are the coercive fields seen in the THE and AHE responses for $\mu_0 H$ parallel to the *ab*-plane and the *c*-axis, respectively. This implies a rotation of the magnetic hard axis of $Fe_{5-x}GeTe_2$ upon exfoliation, from the *c*-axis towards the *ab*-plane, in contrast to what is observed for bulk samples. Experimentally, we find that this reorientation leads to a very large enhancement of both $\rho_{xy}^{\mathrm{A}}$ and $\rho_{xy}^{\mathrm{u,T}}$, with both quantities peaking at a thickness $t \cong 30$ nm. Our micromagnetic simulations indicate that there is a maximum in the topological charge density around this thickness, located between the complete spin homogeneity of very thin films due to an exchange-dominated regime, and the spin inhomogeneity intrinsic to thick films, dominated by the dipolar interactions. We also observe the emergence of a very pronounced hysteresis in the THE, particularly in a temperature range where hysteresis remains completely absent in the longitudinal magnetoresistivity. This implies that the hysteresis observed at higher temperatures is not dominated by the movement and pinning of ferromagnetic domain walls. Instead, we argued that remnant chiral spin textures provide a Hall-like signal even after the external magnetic field is suppressed. This is supported by both our Lorentz transmission electron microscopy (LTEM) measurements that reveal remnant skyrmions upon magnetic field removal, and our micromagnetic simulations indicating that the maximum meron density, and, hence, topological charge-density, peaks at $\mu_0 H = 0$ T. The ensemble of our observations is consistent with intrinsic anomalous and topological Hall responses modulated by the evolution of the spin textures as a function of thickness, temperature, and magnetic field. They also point to the possibility of writing remnant topological spin textures with a magnetic field and electrically detecting them *via* a topological Hall voltage. This exposes the potential of $Fe_{5-x}GeTe_2$ for the development of magnetic memory elements and spintronics in general.

RESULTS/DISCUSSION

Intrinsic versus extrinsic contributions to the anomalous Hall effect in $Fe_{5-x}GeTe_2$

Figure 1 displays the electrical resistivity, $\rho_{xx}$, as a function of the temperature, $T$, for an exfoliated $Fe_{5-x}GeTe_2$ flake of thickness $t$ = 35 nm that was transferred onto pre-patterned Ti:Au electrical contacts and subsequently encapsulated with an exfoliated *h*-BN crystal. $\rho_{xx}$ displays a $T$-dependence and values akin to those seen in bulk samples[11] implying that the exfoliated material is not degraded by the fabrication process, which is performed under inert conditions. A sharp decrease in $\rho_{xx}$ is observed upon cooling below the magneto-structural transition at $T_s$ ~ 110 K[11]. Upon application of a magnetic field (Fig. 1b), $\rho_{xx}(\mu_0 H^c)$ is observed to decrease (negative magnetoresistivity) implying the suppression of spin-scattering processes.

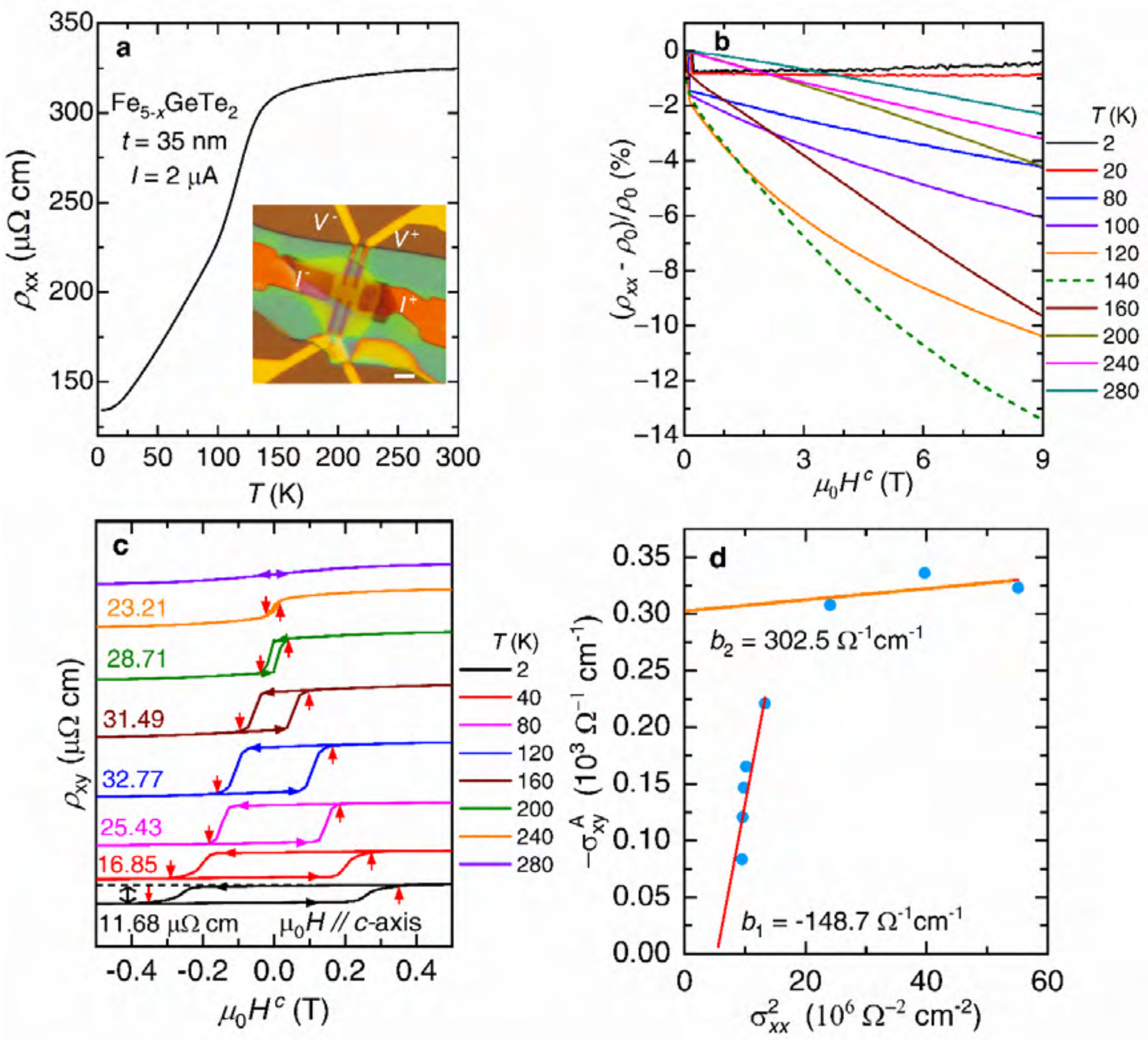

**Figure 1. Hysteresis and intrinsic contribution to the anomalous Hall effect in exfoliated $Fe_{5-x}GeTe_2$. (a) Resistivity ($\rho_{xx}$) as a function of $T$ (K) for a flake of thickness $t$ = 35 nm transferred onto pre-patterned Ti:Au contacts and subsequently encapsulated with a top *h*-BN layer. Inset: Micrograph of the sample, where the scale bar (horizontal white line) indicates a lateral dimension of 5 µm. (b) Magnetoresistivity (($\rho_{xx}$ - $\rho_0$)/$\rho_0$ ) as a function of the magnetic field $\mu_0 H^c$ applied along the *c*-axis for several temperatures for the heterostructure in a. The maximum of the magnetoresistivity is reached at 140 K, as indicated by the dashed green line. (c) Raw Hall resistivity ($\rho_{xy}$) as a function of $\mu_0 H^c$ for several values of $T$. We notice the presence of an AHE response as well as the emergence of a large irreversibility/hysteresis in the Hall response, which is absent in bulk single crystals. Red vertical arrows indicate the values of the coercive field $\mu_0 H_c^c$. (d) Anomalous Hall conductivity ($-\sigma_{xy}^A$) as a function of the square of the conductivity ($\sigma_{xx}^2$). Two regimes were observed resulting in the linear**

**fits $b_1$ (red) and $b_2$ (orange), above and below, respectively, the magneto-structural transition at $T_s$. These coefficients provide the intrinsic contribution to the anomalous Hall response.**

The sharp step seen at low fields results from the emergence of hysteresis, confirming the hard ferromagnetic response[29, 37] of exfoliated $Fe_{5-x}GeTe_2$. The most pronounced negative magnetoresistivity is observed around $T$ = 120 K to 140 K, or just above $T_s$, implying that pronounced spin fluctuations precede the transition at $T_s$. At the lowest $T_s$ and beyond the coercive field $\mu_0 H_c^c$, the magnetoresistivity becomes slightly positive, implying the near total suppression of thermally activated spin fluctuations. The hysteresis and associated coercive field $\mu_0 H_c^c$ (indicated by the vertical red arrows (Fig. 1c)) is more clearly exposed by measurements of the AHE resistivity $\rho_{xy}^{\mathrm{A}}(\mu_0 H^c)$. The hysteresis in the AHE response is reflected by $\rho_{xx}(\mu_0 H^c)$ (Supplementary Figure S1). Note, that in Fig. 1c for fields oriented along the $c$-axis the coercive field for $t$ = 35 nm (~12 unit cells or 36 layers) approaches $\mu_0 H_c^c$ ~ 0.4 T at $T$ = 2 K, with this value being considerably larger than those reported in Ref. [29] for samples having thicknesses ranging from 4 to 15 layers. This suggests that the maximum values of $\mu_0 H_c^c$ are observed in samples having thicknesses ranging between 10- and 20-unit cells and implies thickness-dependent domains, spin textures[34], and magnetic axial anisotropy affecting the ferromagnetic hardness of $Fe_{5-x}GeTe_2$.

Given that both extrinsic and intrinsic mechanisms contribute to the AHE of a ferromagnet, before discussing the thickness dependence of $\rho_{xy}^{\mathrm{A}}$ in $Fe_{5-x}GeTe_2$, one should expose the dominant mechanism. According to the scaling analysis of Ref.[28], the anomalous Hall conductivity follows the empiric relation:

$$\sigma_{xy}^{\mathrm{A}} = -\frac{\rho_{xy}^{\mathrm{A}}}{\left(\rho_{xy}^{\mathrm{A}}\right)^2+\rho_{xx}^2} = -(\alpha\sigma_{xx0}^{-1} + \beta\sigma_{xx0}^{-2})\sigma_{xx}^2 - b \qquad (1)$$

where, $\sigma_{xx0} = 1/\rho_{xx0}$ is the residual conductivity, $\sigma_{xx} = \rho_{xx}/(\rho_{xx}^2 + \rho_{xy}^2)$ is the conductivity, $\alpha$ and $\beta$ are the skew scattering and side-jump scattering terms, respectively, and $b$ is the intrinsic term due to either the Karplus-Luttinger mechanism[30] or the scalar spin chirality. Therefore, as the temperature is varied, $\sigma_{xy}^{\mathrm{A}}$ should scale linearly with $-\sigma_{xx}^2$, with the intercept providing the magnitude of the intrinsic term $b$. A plot of $\sigma_{xy}^{\mathrm{A}}$ as a function of $\sigma_{xx}^2$ (Fig. 1d) reveals two distinct regimes that are linear in $-\sigma_{xx}^2$. These are indicated by linear fits both above (red line) and below (orange line) $T_s$ ~ 110 K that yield very distinct slopes as well as intercepts, or $b$ values. Below $T_s$, one extracts $b_2 \cong 303$ $(\Omega\ \mathrm{cm})^{-1}$ which is a factor of ~ 3 smaller than the value extracted by us for $Fe_3GeTe_2$, *i.e.*, $b$ ~ 940 $(\Omega\ \mathrm{cm})^{-1}$. This last value is close to $b$ ~1100 $(\Omega\ \mathrm{cm})^{-1}$ obtained at room $T$ for Fe films epitaxially grown on undoped GaAs(001)[28]. For this temperature regime, $T < T_s$, the slope $s_2 = (\alpha_2\sigma_{xx0}^{-1} + \beta_2\sigma_{xx0}^{-2})$ yields a value of 5 x $10^{-7}$ $(\Omega\ \mathrm{cm})^{-1} \ll b_2$ indicating that $\sigma_{xy}^{\mathrm{A}}$ is dominated by the intrinsic mechanism. In contrast, for $T > T_s$ the intercept decreases by a factor of ~2 while also changing its sign, such that $b_1$ ~ –150 $(\Omega\mathrm{cm})^{-1}$. This sign change might indicate an electronic reconstruction at $T_s$ or a change in the relative contributions of the Karpus-Luttinger[30] and spin chirality mechanisms. The slope on the other hand, increases by almost two orders of magnitude up to $s_1 \sim 3 \times 10^{-5}$ $(\Omega\ \mathrm{cm})^{-1} \ll |-150\ (\Omega\ \mathrm{cm})^{-1}|$, implying that the intrinsic mechanism still dominates the AHE for $T > T_s$.

It was reported that the spin textures observed in $Fe_{5-x}GeTe_2$, and related phase diagrams are thickness dependent[21, 29, 33] due to a thickness-dependent competition between the relevant exchange interaction(s), the dipole–dipole coupling[33], and a possible reorientation of the perpendicular magnetic anisotropy (PMA). Some of these studies[29, 33] focused on very thin flakes, *i.e.*, from monolayers (1L) to ~15L thick [13, 38-40].

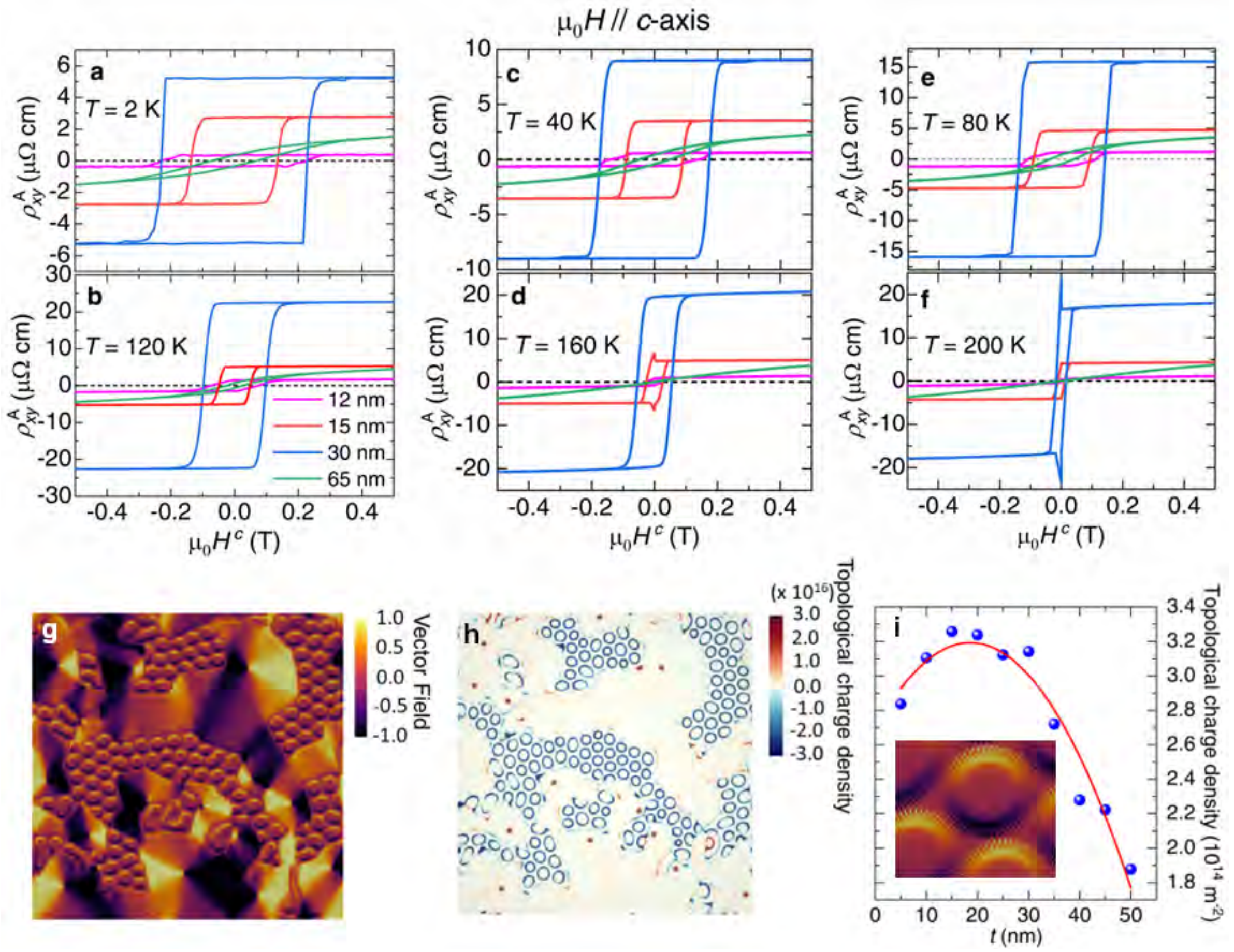

**Figure 2. Anomalous Hall response $\rho_{xy}^{A}$ and micromagnetic simulations. (a-f) $\rho_{xy}^{A}$ as function of $\mu_0 H$ applied along the *c*-axis for six temperatures, 2, 40, 80, 120, 160 and 200 K, respectively. Magenta, red, blue, and green traces correspond to crystal thicknesses of 12, 15, 30 and 65 nm, respectively. Note the increased loop squareness with the increase of the coercive field $\mu_0 H_c^c$ as $T$ is lowered, and the increase in the AHE response as the number of layers, *n*, decreases. The maximum values occurred for an $Fe_{5-x}GeTe_2$ crystal of 30 nm thickness. (g) Snapshot of the magnetization domains obtained *via* micromagnetic simulations at zero field and 0 K, including domains with their magnetization oriented within the conducting planes and domains with the magnetization oriented perpendicular to the planes. As discussed in Refs.[20, 21,36], several domain walls meet at a planar spin vortex that contains a meron at its center. The application of a transverse magnetic field stabilizes skyrmions in domains characterized by an out-of-plane magnetization component[35]. (h) Calculated topological charge density associated with domain structure shown in g. (i) Topological charge density as a function of sample thickness, showing a maximum for 15 nm ≤ $t$ ≤ 25 nm, nearly in agreement with the *t*-dependence of**

**$\rho_{xy}^{A}(t)$. The slight fluctuation in the values of the topological charge (blue dots) are due to several realizations (~10 times) of the random spin seeds that were subsequently averaged for each thickness. The red line is a simple fit to an order-3 polynomial. Inset: magnified image of a simulated magnetic skyrmion.**

Apparently, the Curie temperature $T_c$ remains nearly constant as a function of *t*, decreasing only for the bilayers and monolayers[29]. However, the magnetic phase diagram of $Fe_{5-x}GeTe_2$ measured on much thicker crystals, from 67L to 112L, was still found to be thickness dependent[21], although one would naively expect multilayered crystals to display essentially bulk behavior. To evaluate the effects of manipulating the magnetic phase diagram of $Fe_{5-x}GeTe_2$ *via* exfoliation, we measured the AHE in crystals with thicknesses ranging from 12 nm to 65 nm, or ~12L to ~65L. These crystals are thick enough to remain in the bulk limit, therefore, preserving the previously reported chiral spin textures[20, 36], while being thin enough to affect their phase diagram as a function of the temperature, magnetic-field, and thickness[21].

Thickness dependence of the anomalous Hall response and micromagnetic simulations

For magnetic fields oriented along the c-axis, the anomalous Hall response (Figure 2) is found to display a marked hysteresis, in contrast to what is seen for bulk crystals (Supporting Figure S1), with the coercive field $H_c^c$ increasing as $T$ is lowered, as previously reported in Ref.[29]. The coercivity $H_c^c$ ~ 0.2 – 0.3 mT is a factor of ~2 smaller than the values reported for $Fe_{3-x}GeTe_2$ at low temperatures[41]. Albeit, one still observes $H_c^c \geq 0.1$ T for $T \geq 160$ K, in contrast to $\mu_0 H_c^c$ ~ zero for $Fe_{3-x}GeTe_2$ when $T \geq 135$ K[41]. A maximum in resistivity, $\rho_{xy}^{A,max} \cong 22.6$ μΩ cm, is observed for the anomalous Hall response of the 30 nm thick crystal at $T$ = 120 K which is just above the structural transition at $T_s \cong 110$ K. This value is nearly 5x larger than the one displayed by the $t$ = 65 nm sample and over ~3.6 times larger than $\rho_{xy}^{A,max}$~ 6 μΩ cm extracted from bulk crystals[36]. This enhancement is very difficult to reconcile with extrinsic mechanisms, given that the resistivity, $\rho_{xx}$, as a function of $T$, and, hence, the residual resistivity, $\rho_{xx0}$, of the exfoliated crystals, are identical to those of the bulk crystals (Fig. 1a, and Eq. 1). To put this value in perspective, it is only a factor of 2 smaller than the maximum value of $\rho_{xy}^{\mathrm{A}}$ in $Co_3Sn_2S_2$, a compound claimed to display a giant anomalous Hall response[42] around the same range of temperatures. We are led to conclude that in $Fe_{5-x}GeTe_2$, $\rho_{xy}^{\mathrm{A}}$ is dominated by the intrinsic contribution associated with chiral spin textures, which are modulated by the thickness of the layers. Our data indicates that the maximum values are observed in crystals with thickness ranging from 20 nm to 40 nm. This conclusion is supported by our micromagnetic simulations (see *Methods* for details and Fig. 2g) where we mimicked the ferromagnetic domain structure of exfoliated $Fe_{5-x}GeTe_2$[36]. Some of these domains are characterized by a random in-plane magnetic anisotropy while others display an out-of-plane anisotropy. Our simulation approach resulted in the successful description of the LTEM contrast in $Fe_{5-x}GeTe_2$ in which we observed merons at the converging domain boundaries between planar ferromagnetic domains,[36] in addition skyrmions are stabilized under a perpendicular magnetic field, for domains having an out of plane anisotropy. This leads to a pronounced density of topological charges associated with the merons and skyrmions (Fig. 2h). It turns out that the topological charge density inherent to these spin textures is strongly modulated by the sample thickness, displaying the largest density for thicknesses ranging between 15 and 30 nm in decent agreement with our experimental observations (Fig. 2i). Exfoliation favors the dipolar interaction which leads to a higher degree of order among spin textures. It also favors skyrmions of smaller radii. Both effects conspire to increase the topological charge density within this range of thicknesses. Notice that all crystal thicknesses display a maximum in the anomalous and topological Hall responses around $T \sim 120$ K, suggesting a maximum topological charge density around this temperature, regardless of sample thickness.

The unconventional topological Hall effect

Upon rotating $\mu_0 H$ from the $c$-axis towards the $ab$-plane, one observes a sharp increase in the coercive field, $\mu_0 H_c$ (Figs. 3a-3b), by over one order of magnitude, for an $Fe_{5-x}GeTe_2$ crystal with thickness $t$ = 15 nm. This increase i) indicates that the magnetic easy axis in exfoliated crystals is no longer in the $ab$-plane as is the case for bulk $Fe_{5-x}GeTe_2$ single-crystals[36], and ii) that it surpasses, by a factor >2, the values of $\mu_0 H_c^{ab}$ for $Fe_{3-x}GeTe_2$ whose magnetic hard axis remains within the $ab$-plane in exfoliated crystals[41]. Notice that the large values of $\mu_0 H_c^{ab}$ in exfoliated $Fe_{5-x}GeTe_2$, relative to those of exfoliated $Fe_{3-x}GeTe_2$ crystals, cannot be attributed to a higher degree of disorder that would pin magnetic domain walls.

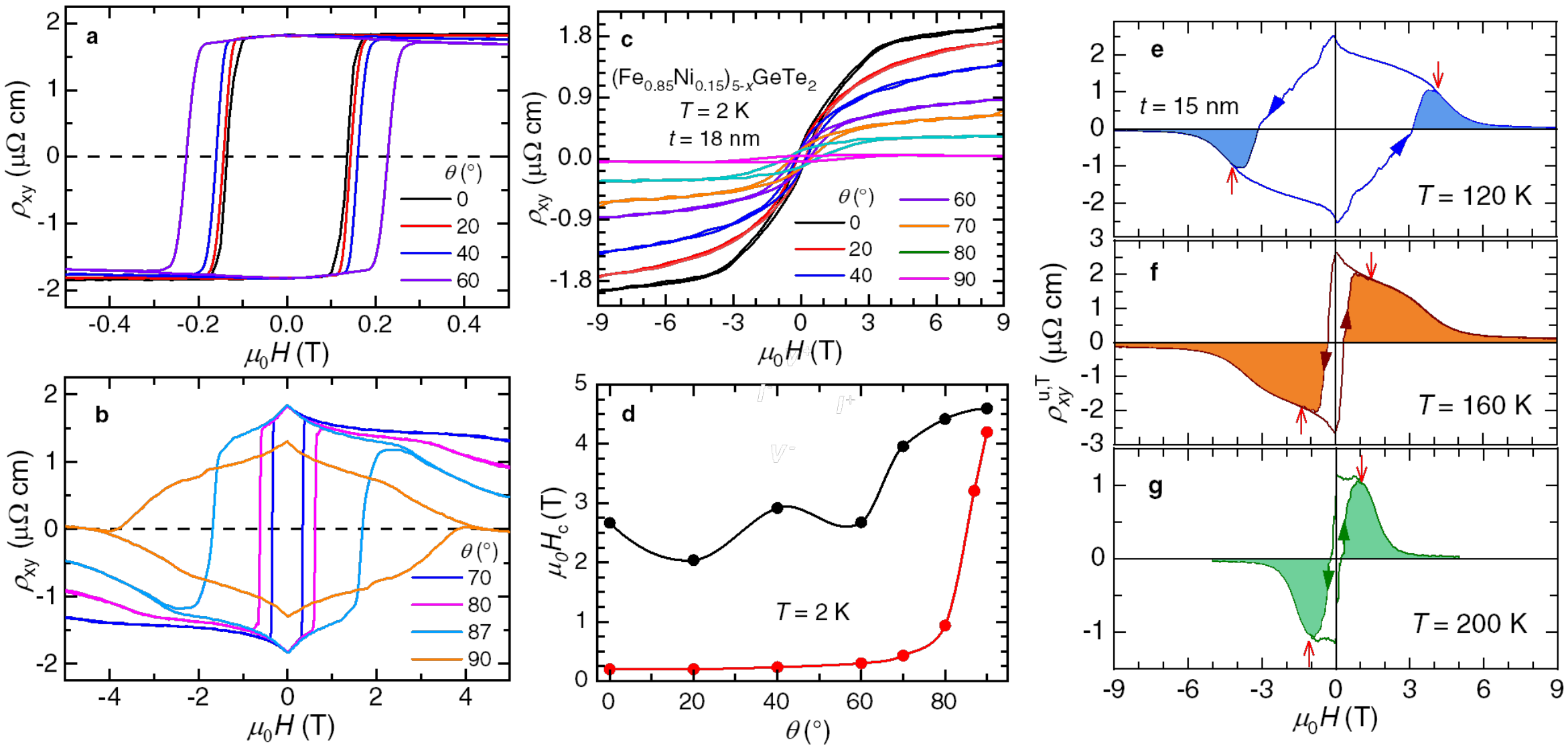

**Figure 3. Coercivity as a function of magnetic field orientation and unconventional THE. (a-b) Raw Hall resistivity $\rho_{xy}$ at $T$ = 2 K as a function of the angle $\theta$ between $\mu_0 H$ and the $c$-axis of a $t$ = 15 nm thick $Fe_{5-x}GeTe_2$ crystal. We observed an increase by more than one order of magnitude in the coercive field $H_c$ as $\mu_0 H$ is rotated towards the $ab$-plane. (c) Raw $\rho_{xy}$ measured from a Ni-doped $Fe_{5-x}GeTe_2$ crystal (composition $(Fe_{0.85}Ni_{0.15})_{5-x}GeTe_2$) of thickness $t$ = 18 nm. Note the similar values of the saturating anomalous Hall response with respect to the value for the $t$ = 15 nm thick $Fe_{5-x}GeTe_2$ crystal. In contrast, $(Fe_{0.85}Ni_{0.15})_{5-x}GeTe_2$ displays one order of magnitude larger values for $\mu_0 H_c^c$, probably due to the disorder inherent to its alloy character. (d) Coercive**

**fields, $\mu_0 H_c$, as a function of the angle $\theta$ between $\mu_0 H$ and the interlayer *c*-axis for both $(Fe_{0.85}Ni_{0.15})_{5-x}GeTe_2$ ($t$ = 18 nm, black markers) and $Fe_{5-x}GeTe_2$ ($t$ = 15 nm, red markers) at $T$ = 2 K. For both compounds, $H_c$ increases rapidly as $\theta$ surpasses 70º, but for both samples it displays similar values around $\theta$ = 90º. (e), (f) and (g) Unconventional THE response $\rho_{xy}^{\mathrm{u,T}}$ resulting from an unconventional measurement geometry, namely magnetic field parallel to the electrical current using a Hall geometry for the voltage leads, and for three temperatures $T$ = 120 K, 160 K, and 200 K, respectively. We note extremely large coercive fields $\mu_0 H_c^{ab}$ (indicated by red arrows) that are over one order of magnitude larger than $\mu_0 H_c^{c}$ at the same temperatures, and the broad peak in $\rho_{xy}^{\mathrm{u,T}}$ for fields beyond $\mu_0 H_c^{ab}$.**

This point is illustrated by the Hall resistivity of an exfoliated Ni-doped $Fe_{5-x}GeTe_2$ single-crystal ($t$ = 18 nm, Fig. 3c), with composition $(Fe_{0.85}Ni_{0.15})_{5-x}GeTe_2$. Details concerning single crystal X-ray diffraction are provided by Table 1 in Methods, Supporting Table S1, and Supporting Figure S3, indicating a homogeneous solid solution whose crystallographic structure can be well refined. Ni doping was found to increase the Curie temperature of $Fe_{5-x}GeTe_2$ although the exact mechanism is not understood[12]. The authors of Ref.[12] mention a few possible mechanisms such as site occupancy, structural modifications that alter the relevant exchange couplings, and the electron doping. For $(Fe_{0.85}Ni_{0.15})_{5-x}GeTe_2$, and for $\mu_0 H$ parallel to the *c*-axis, one observes a marked increase in the magnetic field required to saturate $\rho_{xy}^{\mathrm{A}}(\mu_0 H, \theta)$. Therefore, Ni doping also increases $\mu_0 H_c^c(\mu_0 H, \theta)$, to a value that now exceeds the corresponding ones for $Fe_{5-x}GeTe_2$ by one order of magnitude, due to structural disorder. However, for fields close to the *ab*-plane, the Ni-doped crystal displays almost the same values of $\mu_0 H_c^{ab}(\theta = 90°)$ as the undoped compound. This implies that structural disorder, expected to pin the movement of magnetic domain walls upon sweeping a magnetic field, has little influence on the pronounced hysteresis superimposed onto $\rho_{xy}^{\mathrm{u,T}}$, observed when $\mu_0 H \parallel j \parallel ab$-plane. The temperature dependence of $\rho_{xy}^{\mathrm{u,T}}$ (Figs. 3e-3g) reveals not only an important increase in the irreversibility field $\mu_0 H_c^{ab}(\theta = 90°)$ (vertical red arrows) as $T$ is lowered, but a broad maximum whose amplitude and field width (color shaded areas) is $T$-dependent. Here, we emphasize that i) great care was taken to carefully align the field along the direction of the electrical current, and ii) the behavior of $\rho_{xy}^{\mathrm{u,T}}(\mu_0 H)$, namely the presence of a broad peak, bears no resemblance to the behavior of either the magnetization (Figure S1) or the longitudinal magnetoresistivity (Figure S4). For example, neither the magnetization, nor the longitudinal magnetoresistivity, displays a zero value above $\mu_0 H$ ~ 6 T at $T$ = 200 K. This indicates that $\rho_{xy}^{\mathrm{u,T}}(\mu_0 H)$ cannot be attributed to an AHE component due to simple sample misalignment. This is further confirmed by data collected on crystals having $t$ = 15 nm and 30 nm, which displays a very small $\rho_{xy}^{\mathrm{u,T}}(\mu_0 H)$ signal at $T$ = 240 K, while still exhibiting a sizeable AHE response (Figure S5). Instead, as argued in Refs.[32, 36], this corresponds to an unconventional THE response (*i.e.*, unconventional measurement geometry) resulting from magnetic field-induced chiral spin textures that bend the electronic orbits and induce a Hall-like response. Our results imply that these textures are affected not only by temperature, and, hence, the magneto-structural transition at $T_s$ ~ 110 K, but also by thickness and/or the geometry of the exfoliated crystals that affects the magnetic anisotropy and the relative strength between dipolar and exchange interactions. In contrast, point disorder is clearly detrimental to $\rho_{xy}^{\mathrm{u,T}}$ as can be seen in Fig. 3c for $(Fe_{0.85}Ni_{0.15})_{5-x}GeTe_2$ which reveals no evidence of the unconventional topological Hall response for fields nearly along the *ab*-plane. Albeit, this compound still exhibits basically the same maximum amplitude for the anomalous Hall response relative to the undoped compound.

Remnant unconventional topological Hall response and Lorentz TEM images

To illustrate this point, we carefully measured $\rho_{xy}^{\mathrm{u,T}}$ in four exfoliated crystals, with thicknesses of $t$ = 12 nm, 15 nm, 30 nm, and 50 nm, as a function of $\mu_0 H \parallel j \parallel ab$-plane, for several temperature values (Figure 4). All samples display a very pronounced, anti-symmetric, and hysteretic *T*-dependent signal that leads to coercive fields as large as $\mu_0 H_c^{ab}$ ~ 5.65 T, in contrast to $\mu_0 H_c^{c}$ ~ 0.2 T (for the *t* = 30 nm thick sample at *T* = 40 K).

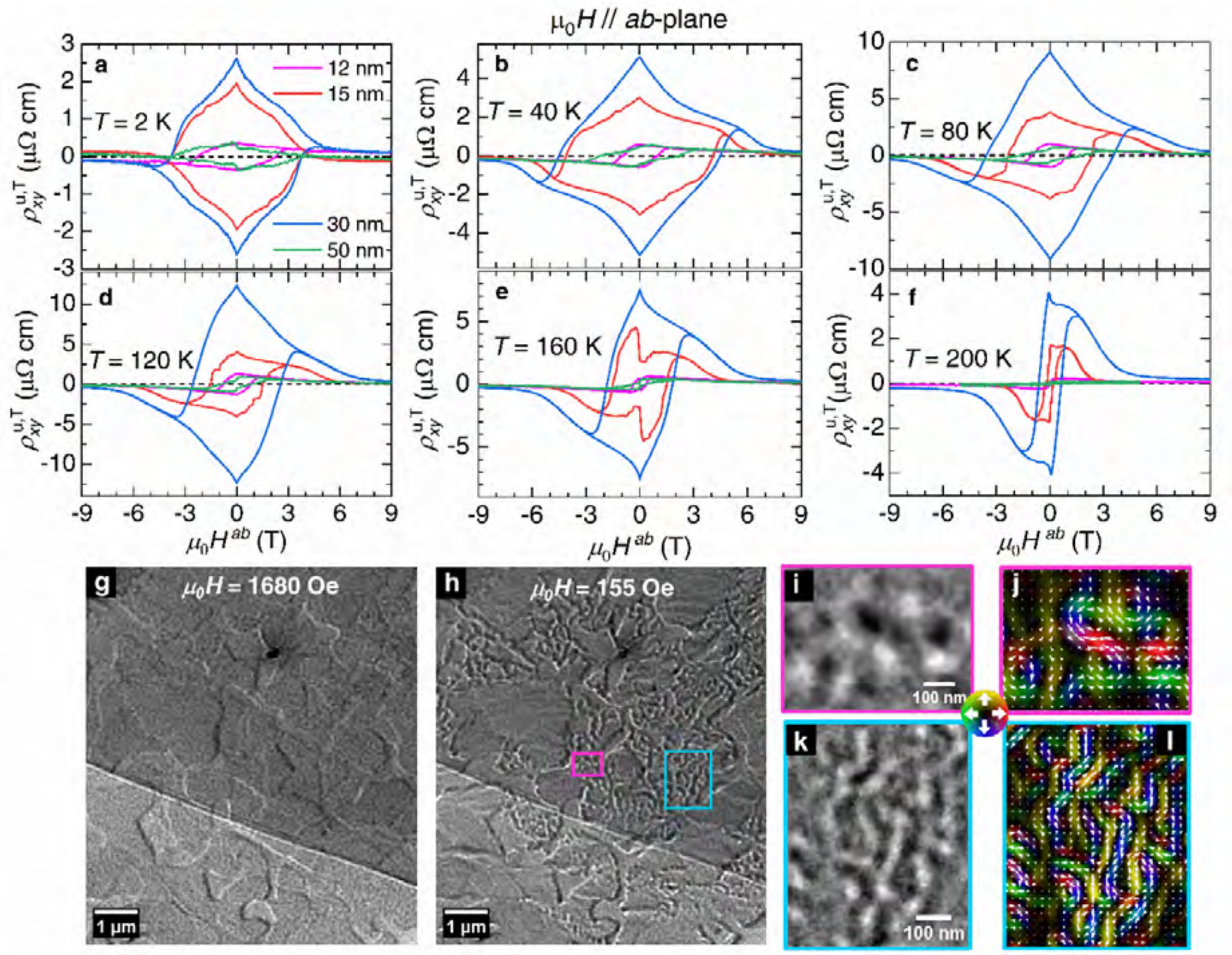

**Figure 4. Hysteretic Hall-like response along the planar direction and remnant chiral spin textures. (a-f) Unconventional THE response for four samples having different thickness, *t*, as a function of $\mu_0 H$ aligned parallel to the electrical current flowing within the *ab*-plane, for *T* = 2 K, 40 K, 80 K, 120 K, 160 K, and 200 K, respectively. Magenta, red, blue, and green traces correspond to samples with *t* = 12 nm, 15 nm, 30 nm, and 50 nm, respectively. Notice the increase in $\mu_0 H_c^{ab}$ as *T* decreases, and the observation of a broad peak beyond $\mu_0 H_c^{ab}$ (see, Figs. 3e – 3g) which displays a maximum value for 120 K ≤ *T* ≤ 160 K. Both the reversible and irreversible components of this unconventional Hall response are largest for the sample with thickness *t* = 30 nm. (g) LTEM image of an exfoliated $Fe_{5-x}GeTe_2$ crystal encapsulated with a graphene layer (darker region) in**

**an out-of-plane field of $\mu_0 H$ = 1680 Oe. Under this field, the mgnetic domains are polarized. (h) LTEM image of the same region in the remnant field out-of-plane field of the objective lens ($\mu_0 H$ ~ 155 Oe), revealing labyrinthine domains (area enclosed by blue rectangle) as well as skyrmions (magenta rectangle). To image Néel-like out-of-plane domains, for both panels g and h, the sample was tilted by 25° about the *x*-axis. (i-j) Magnified image and magnetic induction map, reconstructed using the transport of intensity equation[43] respectively, of the area enclosed by the magenta rectangle. This reveals the projected component of magnetic induction of the inner and outer regions in Néel skyrmions, showing a bound vortex / anti-vortex structure. Colors correspond to the in-plane orientation of the local magnetization. (k-l) Labyrinthine domains and corresponding magnetic induction maps for the area enclosed by the blue rectangle.**

Most importantly, this hysteresis leads to pronounced *T*- and field-sweep-orientation-dependent remnant values for $\rho_{xy}^{\mathrm{u,T}}$ ($\mu_0 H = 0$ T) or $\rho_{xy}^{\mathrm{u,T,rem}} \cong 12.5$ μΩ cm for the *t* = 30 nm sample at 120 K. This hysteresis in $\rho_{xy}^{\mathrm{u,T}}$ it is not observed in the longitudinal magnetoresistivity (Figure S4) for *T* > 100 K. Furthermore, for all temperatures the longitudinal magnetoresistivity always returns to its original zero-field value after sweeping the magnetic field along either direction. This implies that ferromagnetic domain wall pinning, and field-induced domain wall reconfiguration, are not the main mechanisms leading to hysteresis. Domain walls are barriers for carrier motion and are reported to affect the resistivity in many systems[44-48] leading to hysteresis. In contrast, and as seen through Figs. 4 to 4f, $\rho_{\mathrm{xy}}^{\mathrm{u,T}}$ increases continuously as the field is reduced to zero, implying that the magnetic structures responsible for it are not pinned, as conventional FM domain walls, but evolve continuously as a function of decreasing magnetic field. Therefore, the hysteresis seen solely in $\rho_{xy}^{\mathrm{u,T}}$, and leading to a finite Hall-like response at zero magnetic field, ought to result from remnant magnetic field-induced chiral spin textures, such as merons, upon field removal. This assertion is supported by our LTEM results on exfoliated samples of $Fe_{5-x}GeTe_2$ (Figs. 4g-4l). LTEM is a powerful magnetic domain visualization technique with sub-micron resolution, or higher resolution then magneto-optical Kerr effect, being comparable to near field probes[49], although being sensitive to the component of the magnetization perpendicular to the electron beam. We observed that spin textures (*e.g.*, skyrmions and merons) remain after the external magnetic field has increased to the point of completely removing their presence and has been subsequently reduced to the lowest values attainable within the LTEM instrument (Fig. 4h). Our micromagnetic simulations indicate that the largest density of magnetic textures, and, hence, largest topological charge density is observed at $\mu_0 H$ = 0 T (Figure S6). An increasing in-plane field suppresses the spin textures, especially merons, as the exchange energy forces them to transition to saturated spin configurations (stripe domains). In Figure 3, one can see that the broad reversible peaks observed in $\rho_{xy}^{\mathrm{u,T}}$ beyond the irreversibility field $H_c^{ab}$, display maximum values in the temperature range ~80 K to ~160 K, precisely in the temperature range where the hysteresis becomes most pronounced (despite its absence in the longitudinal MR), leading to the largest values of $\rho_{xy}^{\mathrm{u,T,rem}}$ and, thus, implying a correlation with $\rho_{xy}^{\mathrm{u,T}}$. Note that a Hall effect at zero field is not the exclusive purview of the quantum AHE[50-52], but has also been observed in a compound characterized by chiral, albeit, dynamical spin-ice configurations[53, 54]. Note also that the values of $\rho_{xy}^{\mathrm{u,T,max}}(t$ = 30 nm) approach the maximum values of the colossal THE[55] observed in $MnBi_4Te_7$, while $\rho_{xy}^{\mathrm{u,T,rem}}$ surpasses them. $\rho_{xy}^{\mathrm{u,T,rem}}$ cannot be ascribed to defects, since it is difficult to understand how defects could lead to behavior that is field orientation dependent or would lead to well oriented magnetic moments at zero field (*i.e.*, mimicking ferromagnetism) as to provide a global Hall response at zero field. Finally, we point out that $\rho_{\mathrm{xy}}^{\mathrm{u,T}}(\mu_0 H)$ changes its sign at either side of $\theta$ = 90$^{\mathrm{o}}$, suggesting that spin textures are written by the out-of-the-plane component of the magnetic field (Figure S7). We found that it is quite difficult to align the field very carefully within the *ab*- plane (*i.e.*, with a misalignment inferior to $\Delta\theta \sim 0.05^{\circ}$) and in this way detect $\rho_{\mathrm{xy}}^{\mathrm{u,T}}(\mu_0 H)$ = 0 μΩ cm. Our micromagnetic simulations[36, 56, 57] mimicking an exfoliated sample (Fig. S8), yields a topological charge density as a function of the interlayer magnetic field which displays a striking resemblance with the topological Hall response shown in Fig. 4. The Hall resistance $R_{\mathrm{xy}}$ is proportional to the topological number $N_{Sk}$ via $R_{\mathrm{xy}} \propto \int_{-y_0}^{y_0}\int_{-x_0}^{x_0} N_{Sk}(x-x', y-y')dx'dy'$ [58], which is related to the emergent field created by the spin textures via $\boldsymbol{B}_{em} \propto N_{Sk}\hat{e}_z$ [59]. Therefore, the topological Hall resistance is directly proportional to the emergent field induced by the spin textures and associated topological charge density due to the presence of skyrmions and merons.

We provide a summary of the irreversible fields as well as the amplitudes for both the AHE and the unconventional THE responses in Figure 5. The data for four sample thicknesses that are well beyond the monolayer limit suggest that thicker crystals display larger values of $H_c^c$s for magnetic fields applied perpendicular to the layers (Fig. 5a) than thinner crystals[29]. As for $H_c^{ab}$, seemingly the thinnest flakes display the smallest values (Fig. 5b), suggesting the evolution of the spin textures as a function of *t*[33, 34]. However, both $\rho_{xy}^{A}$ and $\rho_{xy}^{\mathrm{u,T}}$ display a pronounced thickness dependence despite the multilayered nature of the measured crystals (Fig. 5c-5d). Given that $\rho_{xy}^{\mathrm{A}}$ in $Fe_{5-x}GeTe_2$ is dominated by the intrinsic contribution, as shown in Fig. 1 and related discussion, thickness-dependent values for $\rho_{xy}^{\mathrm{A}}$ indeed support the notion of spin textures evolving as a function of sample thickness. For instance, in Fig. 5c the maximum values of $\rho_{xy}^{\mathrm{A}}$, observed at $T \approx 130$ K for both the $t$ = 30 nm and $t$ = 12 nm samples, are 22.6 μΩ cm and 1.76 μΩ cm, respectively. This is a broad range of values that contrasts markedly with the reproducibility of the bulk value[20, 36], *i.e.*, $\rho_{xy}^{A} \sim$ 4.5 to 6.3 μΩ cm.

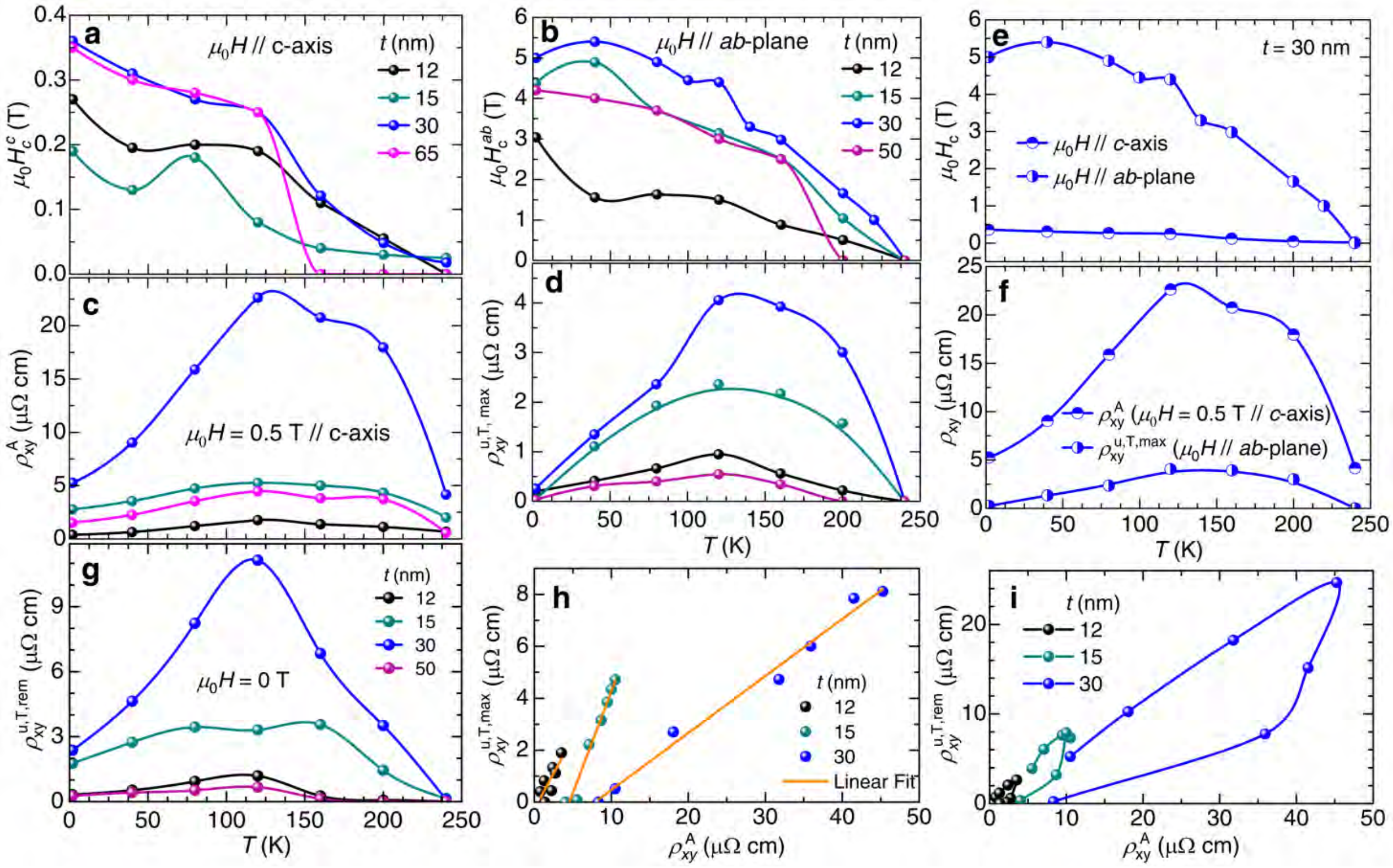

**Figure 5. Correlating coercive fields, anomalous and unconventional topological Hall signals as a function of thickness. (a-b) Coercivity fields $\mu_0 H_c^c$ and $\mu_0 H_c^{ab}$, respectively, as functions of temperature (*T*) for five samples with thicknesses of 12 nm (black), 15 nm (green), 30 nm (blue), 50 nm (purple), and 65 nm (magenta). (c) Anomalous Hall resistivity $\rho_{xy}^{A}$ under $\mu_0 H$ = 0.5 T applied along the *c*-axis. For all thicknesses, $\rho_{xy}^{A}$ increases as *T* decreases, reaching a maximum value just above the magnetostructural transition at $T_s$ ~ 110 K. (d) Amplitude of the maximum $\rho_{xy}^{u,T,max}$ observed beyond $\mu_0 H_c^{ab}$ in the antisymmetric planar Hall-like response of topological character. For all four samples $\rho_{xy}^{u,T,max}$ peaks just above $T_s$, with the *t* = 30 nm sample displaying the most pronounced response over the entire *T* range. e, $H_c^c$ and $H_c^{ab}$ as functions of *T* for the *t* = 30 nm sample. (f) Comparison between the temperature dependences of $\rho_{xy}^{A}(\mu_0 H = 0.5\ \mathrm{T})$ and $\rho_{xy}^{u,T,max}$ for the *t* = 30 nm sample, revealing that both variables peak around $T_s$ albeit decreasing as *T* is lowered. (g) Remnant value of the unconventional topological Hall response $\rho_{xy}^{u,T,rem} = \rho_{xy}^{u,T}(\mu_0 H = 0\ \mathrm{T})$ as a function of *T* and for all four sample thicknesses. Note that $\rho_{xy}^{u,T,rem}$ also displays a maximum around the same temperature (*T* ~ 120 K) where maxima are observed in both $\rho_{xy}^{A}$ and $\rho_{xy}^{u,T}$. (h) Amplitude of the peak $\rho_{xy}^{u,T,max}$ observed in $\rho_{xy}^{u,T}$ beyond $\mu_0 H_c^{ab}$**

**as a function of $\rho_{xy}^{\mathrm{A}}$. One observes a linear relation (orange lines are linear fits) between both quantities, despite the different field orientations used to measure each variable, implying that both are driven by the same underlying physics. i, $\rho_{xy}^{\mathrm{u,T,rem}}$ as a function of $\rho_{xy}^{\mathrm{A}}$ for the different samples revealing a nearly linear relation until the magneto-structural transition is reached, from which point $\rho_{xy}^{\mathrm{u,T,rem}}$ decreases steeply.**

Two additional and important aspects should be considered: Firstly, the suppression of both $\rho_{xy}^{\mathrm{A}}$ and $\rho_{xy}^{\mathrm{u,T}}$ (Figs. 5c-5d) upon cooling below the magnetostructural transition at $T_\mathrm{s}$ ~ 110 K, with the former remaining finite and the latter approaching zero at the lowest temperatures. Secondly, the scaling of $\rho_{xy}^{\mathrm{u,T}}$ with $\rho_{xy}^{\mathrm{A}}$ suggesting that $\rho_{xy}^{\mathrm{u,T}}$is also driven by the intrinsic contribution. This is supported by the behavior of $\rho_{xy}^{\mathrm{u,T,rem}}$ as a function of $t$ and $T$ (Fig. 5g), which mimics that of $\rho_{xy}^{\mathrm{A}}(t,T)$, and the linear scaling between $\rho_{xy}^{\mathrm{u,T.max}}$ and $\rho_{xy}^{\mathrm{A}}(t,T)$ (Fig. 5h) despite the orthogonality in field orientations used to measure the two quantities. We also observe an almost linear relation between $\rho_{xy}^{\mathrm{u,T,rem}}$ and $\rho_{xy}^{\mathrm{A}}(t,T)$ for $T > T_\mathrm{s}$ implying that the intrinsic, topological charge dominated mechanism, leads to the sizeable values of $\rho_{xy}^{\mathrm{u,T,rem}}$ (Fig. 5i). Finally, the samples displaying the largest values for both $\rho_{xy}^{\mathrm{u,T,max}}$ and $\rho_{xy}^{\mathrm{A}}$ are also the ones exhibiting the largest coercive fields $\mu_0 H_c^c$ and $\mu_0 H_c^{ab}$ (Figs. 5e-5f for $t$ = 30 nm). This confirms that the relative change in the values of the exchange constants, magnetic axial anisotropy, and dipolar interactions favors more robust chiral textures with respect to the application of an external magnetic field. These chiral spin textures should affect the Berry phase of the charge carriers in both real and reciprocal spaces, to yield the anomalous and unconventional topological Hall responses[27] as reported here. Although the spin textures for magnetic fields applied along the *c*-axis have been studied *via* LTEM[20, 21, 33, 36], their presence and evolution as a function of temperature and fields applied in plane remains to be investigated. Notice that an analysis of the conventional THE as a function of both the magnetization and magnetoresistivity, is not possible in exfoliated flakes, given the impossibility of accurately measuring their absolute magnetization. In contrast, the unconventional THE is not masked by a superimposed AHE component and conveniently exposing solely the topological contribution.

CONCLUSIONS

In exfoliated $Fe_{5-x}GeTe_2$, in-plane coercive fields exceed, by more than one order of magnitude, those for fields applied perpendicularly to the planes, as well as those reported for its sister compound $Fe_{3-x}GeTe_2$. In relatively thick multilayered flakes that would be expected to display bulk behavior, both the anomalous Hall and unconventional topological Hall responses are strongly thickness dependent. This suggests that both Hall signals have a similar origin, *i.e.*, the exfoliation tuning of the relative strengths of dipolar and exchange interactions. This modulation affects the stability and, hence, the density of chiral spin textures affecting the Berry phase of the moving charge carriers in real and reciprocal spaces. This is supported by a scaling analysis of the anomalous Hall-effect of $Fe_{5-x}GeTe_2$ indicating that it is dominated by the intrinsic Berry phase contribution. The unconventional THE displays a sizeable remnant value at zero-field and at temperatures for which the magnetoresistivity displays no hysteresis. This indicates that the hysteresis, and the finite zero-field Hall-like response, result from remnant, field-induced chiral spin textures with a well-defined dominant chirality, or topological charge. This is supported by our LTEM study in $Fe_{5-x}GeTe_2$ and the micromagnetic simulations which reveal the suppression of planar ferromagnetic domains upon increasing the in-plane magnetic field, but the remanence of skyrmions upon field removal. The largest density of merons, at the vertex between planar ferromagnetic domains, is found precisely at zero field. Hysteresis and remanence results from a stable imbalance between positively- and negatively-topologically-charged spin textures leading to a hitherto unreported topological Hall effect at zero field, which is strongly thickness dependent. Several important aspects of our study remain to be understood, like to the suppression of both anomalous and topological Hall responses upon reducing the temperature below $T$ ~ 120 K. This will require a detailed study on the evolution of the spin textures as a function of $T$ < 120 K. Another quite intriguing aspect, is the observation of a pronounced topological Hall response under planar fields in the order of several tesla, since all spin textures have been suppressed when similarly intense fields were applied along the *c*-axis. Intense magnetic fields pose a major technical challenge for available magnetic domain imaging techniques. The important aspect is that our observations indicate that exfoliated $Fe_{5-x}GeTe_2$ provides a platform for the writing and electrical detection of topological spin textures aiming at energy-efficient devices based on vdW ferromagnets. In effect, topological spin textures have been proposed for the development of several computational schemes, ranging from neuromorphic[60], probabilistic[61], temporal[62], reconfigurable[63], high density reservoir computing[64] and even quantum computing[65, 66]. They were also proposed for the development of memory elements[67, 68], such as skyrmion race tracks[69]. Some of these schemes rely on magnetic tunnel junctions to detect the presence of skymions and/or current pulses to write/delete them. Here, we showed that their presence can be detected *via* a simple Hall response, even in absence of an external magnetic field, or that a modest external field, and in particular its orientation, can be used to write specific topological spin textures in thin van der Waals ferromagnets that can be grown in large area[70]. Furthermore, remnant spin textures, in absence of an external magnetic field, provide a potential for the development of non-volatile information carriers, whose spin textures can be controlled via global or local magnetic fields. This, coupled with the simple methods described here to write and detect skymions, unveils the potential of van der Waals ferromagnets for applications. It remains to be seen if this effect can be brought to room temperature *via* Ni or Co doping, which is known to increase its Curie temperature.

## METHODS/EXPERIMENTAL

**Single-crystal synthesis.** Single crystals of $Fe_{5-x}GeTe_2$ were synthesized through a chemical vapor transport technique. Starting molar ratios of 6.2:1:2 for Fe, Ge, and Te, respectively, were loaded into an evacuated quartz ampoule with approximately 100 mg of $I_2$ acting as the transport agent. After the initial warming, a temperature gradient of 75°C was established between a 775 °C and 700 °C zone of a 2-zone furnace and maintained for 14 days, during which large single crystals nucleated at the 700°C zone. Samples were subsequently quenched in ice water to yield the maximum Curie temperature[11]. Crystals used in this paper are from the same batch used in previous experiments[71]. Crystals were washed in acetone and subsequently isopropyl alcohol to remove residual iodine from their surface. According to Energy Dispersive Spectroscopy, the values of *x* are found to range between 0.15 and 0.

**Single-Crystal X-ray diffraction**. The crystal structure of Ni doped $Fe_{5-x}GeTe_2$ was determined using a single crystal fragment on a Bruker D8 Quest Kappa single crystal X-ray diffractometer equipped with an I$\mu$S microfocus source (Mo $K_{\alpha}$, $\lambda$ = 0.71073 Å), HELIOS optics monochromator, and PHOTON III CPAD detector. Diffraction data was integrated using the Bruker SAINT program and an absorption correction was applied to the intensities with a multi-scan method in SADABS 2016/2. A preliminary starting model was obtained using the intrinsic phasing method in SHELXT. Our best fit model resulted in the crystallographic structure $R\bar{3}m$ and cell dimensions of $a$ = 4.037(4) Å, and $c$ = 29.11(4) Å; additional reflections were observed in the $hk$1plane and were not indexed in final model. As Ni and Fe are indistinguishable *via* X-ray diffraction, energy dispersive spectroscopy values of $Fe_{3.95}Ni_{0.67}GeTe_2$ were used to constrain structure occupation factors of Fe and Ni yielding a composition of $Fe_{4.01}Ni_{0.68}GeTe_2$.

**Table 1| Crystallographic Data and Refinement Parameters of $Fe_{4.01}Ni_{0.68}GeTe_2$**

| Empirical Formula | $Fe_{4.01}Ni_{0.68}GeTe_2$ |
|---|---|
| Space group, crystal system | $R\bar{3}m$:H |
| Lattice Parameters | |
| *a* (Å) | 4.037 (4) |
| *c* (Å) | 29.11 (4) |
| *Volume* [$Å^3$] | 410.8(1) |
| *Z* | 1 |
| Density [$g/cm^3$] | 7.177 |
| Absorption coefficient [$mm^{-1}$] | 28.34 |
| F(000) | 778 |
| Crystal size [$mm^3$] | 0.05 × 0.04 × 0.02 |
| $\theta$ range [°] | 2.1–28.3 |
| Index range | |
| *h* | −5→5 |
| *k* | −5→5 |
| *l* | −37→37 |
| Number of reflections | 3059 |
| Unique reflections | 166 |
| Parameters/restraints | 17/0 |
| $R_{int}$ | 0.068 |
| $\Delta\rho_{max/min}$ | 1.95/-2.27 |
| *GoF* | 1.24 |
| R [$F^2 > 2\sigma(F^2)$] | 0.036 |
| $wR_2$ ($F^2$) | 0.085 |

$R = \sum|(|F_o|-|F_c|)|/\sum |F_o|$ and $wR_2 = \{\sum w[(F_o)^2 - (F_c)^2 ]^2/\sum w [(F_o)^2]^2\}^{1/2}$

**Table S1** and **Figure S3** in the Slist atomic coordinates and display Bragg reflections used for the X-ray refinement of the crystallographic structure, respectively.

**Electrical transport measurements.** To prevent oxidation, single crystals were exfoliated under argon atmosphere, within a glove box containing less than 10 parts per billion in oxygen, and water vapor. These were subsequently dry transferred onto Ti:Au contacts pre-patterned on a $SiO_2$/*p*-Si wafer using a polydimethylsiloxane stamp, and subsequently encapsulated among *h*-BN layers, with both operations performed under inert conditions. Titanium and gold layers were deposited *via* e-beam evaporation techniques, and electrical contacts fabricated through electron beam lithography. All measurements were performed in a Quantum Design Physical Property Measurement System.

**Cryogenic Lorentz transmission electron microscopy**. Single crystalline $Fe_{5-x}GeTe_2$ was mechanically exfoliated directly onto homemade polydimethylsiloxane stamp inside an argon filled glovebox. Prior to its utilization, the stamp was rinsed in acetone and isopropyl alcohol to clean its surface. After appropriate crystal thicknesses and dimensions were identified *via* optical contrast, the selected crystal(s) was transferred onto a window of a silicon-nitride based transmission electron microscopy grid. Few-layer graphite (14 nm thick) was transferred onto the $Fe_{5-x}GeTe_2$ flake through the same dry transfer method to protect the sample from oxidization. To characterize the magnetic domains, the out-of-focus LTEM images were taken in a JEOL 2100F TEM operating in Lorentz mode (Low Mag), a perpendicular magnetic field aligned parallel to electron beams being generated by applying a small amount of current to the objective lens. The magnetic induction maps were reconstructed based on transport-of-intensity equation (TIE) method using the PyLorentz software package[43].

**Micromagnetic Simulations.** The numerical simulations were performed using the Mumax3 solver[72], which allowed larger simulation sizes relatively to atomistic spin dynamics[57, 73-8269-80]. Regions of in-plane and out-of-plane anisotropy comprised of Thiessen polygons were generated with Voronoi tessellation, using a grain size of 200 nm. The material parameters were chosen to be[19] exchange constant $A$ = 10 pJ $\mathrm{m}^{-1}$, Gilbert damping α = 0.3, saturation magnetisation $M_s$ = 630 kA $\mathrm{m}^{-1}$ along *c*-axis, $M_s$ = 730 kA $\mathrm{m}^{-1}$ along *ab*-plane, $D$ = 1.2 mJ $\mathrm{m}^{-2}$ and $K_u$ = 2.5 kJ $\mathrm{m}^{-3}$ in the out-of-plane regions, and periodic boundary conditions were applied in the lateral film dimensions. The magnetization was initially randomized before relaxing the magnetic material in the presence of a $+\hat{z}$ directed magnetic field, before being reduced to zero-field. For hysteresis calculations, the external field was applied in-plane and varied in steps of 10 Oe, and the average topological charge density was recorded as a function of in-plane field at each step. The simulations were performed with a grid resolution of 2 nm to ensure sufficient accuracy in the topological charge calculations.

## ASSOCIATED CONTENT

Supporting Information

Magnetic response of bulk single-crystalline $Fe_{5-x}GeTe_2$, transverse magnetoresistivity $MR$ as a function of the magnetic field $\mu_0 H$ applied along the $c$-axis of a 15 nm thick $Fe_{5-x}GeTe_2$ crystal and for several temperatures, atomic coordinates of $Fe_{4.01}Ni_{0.68}GeTe_2$ from single crystal X-ray diffraction, single crystal X-ray diffraction of $Fe_{4.01}Ni_{0.68}GeTe_2$, longitudinal magnetoresistivity $MR$ ($\mu_0 H \parallel j$ with $j$ being the current density) as a function of the magnetic field $\mu_0 H$ applied along the $c$-axis of a 15 nm thick $Fe_{5-x}GeTe_2$ crystal and for several temperatures, hysteretic response in both the anomalous Hall and unconventional topological Hall response of exfoliated $Fe_{5-x}GeTe_2$ crystals, topological charge density as a function of the magnetic field according to micromagnetic simulations, unconventional topological Hall response for fields very close to the *ab*-plane, calculation of the topological charge as a function of the inter-planar magnetic field, parameters used for the micromagnetic simulations.

The Supporting Information is available free of charge at https://pubs.acs.org/doi/

**Corresponding authors**

**Luis Balicas** - [1]*National High Magnetic Field Laboratory, Tallahassee, FL 32310, USA. 2Department of Physics, Florida State University, Tallahassee, USA. https://orcid.org/0000-0002-5209-0293,* Email: lbalicas@fsu.edu

**Elton J. G. Santos** – *Institute for Condensed Matter and Complex Systems, School of Physics and Astronomy, The University of Edinburgh, EH9 3FD, UK.* [ǂ]*Higgs Centre for Theoretical Physics, The University of Edinburgh, EH9 3FD, UK. https://orcid.org/0000-0001-6065-5787,* Email: esantos@ed.ac.uk

## AUTHOR CONTRIBUTIONS

A.M. and W.Z. fabricated the heterostructures under inert conditions. A.M. performed the transport measurements. Y.L. and C.P. performed the L-TEM study and analysis. B.W.C. and J. M. grew and characterized the $Fe_{5-x}GeTe_2$ bulk single-crystals. C.M. and E.J.G.S. performed the micromagnetic simulations, and the theoretical description of the data. M.B. performed the X-ray diffraction of the $(Fe_{0.85}Ni_{0.15})_{5-x}GeTe_2$ single crystals, and analyzed the results with G.T.M., and J.Y.C. L.B. wrote the manuscript with input from all co-authors. All authors read the manuscript, commented on it, and have given their approval to the final version.

**ACKNOWLEDGEMENTS**

L.B. acknowledges support from the US DoE, BES program through award DE-SC0002613 US (synthesis and measurements), US-NSF-DMR 2219003 (heterostructure fabrication) and the Office Naval Research DURIP Grant 11997003 (stacking under inert conditions). J.Y.C acknowledges NSF DMR-2209804 and the Welch Foundation through AA-2056-20220101. The National High Magnetic Field Laboratory acknowledges support from the US-NSF Cooperative agreement Grant numbers DMR-1644779 and DMR-2128556, and the state of Florida. EJGS acknowledges computational resources through CIRRUS Tier-2 HPC Service (ec131 Cirrus Project) at EPCC (http://www.cirrus.ac.uk) funded by the University of Edinburgh and EPSRC (EP/P020267/1); ARCHER UK National Supercomputing Service (http://www.archer.ac.uk) *via* Project d429. EJGS acknowledges the EPSRC Open Fellowship (EP/T021578/1) and the Edinburgh-Rice Strategic Collaboration Awards for funding support. Work at Argonne (Y.L, C.P) was funded by the US Department of Energy, Office of Science, Office of Basic Energy Sciences, Materials Science and Engineering Division. Use of the Center for Nanoscale Materials, an Office of Science user facility, was supported by the U.S. Department of Energy, Office of Science, Office of Basic Energy Sciences, under Contract No. DE-AC02-06CH11357.

**Supporting information for manuscript titled: "Writing and detecting topological charges in exfoliated $\mathbf{Fe_{5-\mathit{x}}GeTe_2}$"**

*Alex Moon[§,£], Yue Li[¤], Conor McKeever[¥], Brian W. Casas[§], Moises Bravo[ʅ], Wenkai Zheng[§,£], Juan Macy[§,£], Amanda K. Petford-Long[¤,ƺ], Gregory T. McCandless[ʅ], Julia Y. Chan[ʅ], Charudatta Phatak[¤,ƺ], Elton J. G. Santos[¥,ǂ,*], and Luis Balicas[§, £,*]*

[§]*National High Magnetic Field Laboratory, 1800 E. Paul Dirac Dr. Tallahassee, FL 32310, USA.*
[£]*Department of Physics, Florida State University, 77 Chieftan Way, Tallahassee, FL 32306, USA.*
[¤]*Materials Science Division, Argonne National Laboratory, Lemont, 60439, IL, USA.*
[¥]*Institute for Condensed Matter and Complex Systems, School of Physics and Astronomy, The University of Edinburgh, EH9 3FD, UK.*
[ʅ]*Department of Chemistry and Biochemistry, Baylor University, Waco, TX 76798, USA.*
[ƺ]*Department of Materials Science and Engineering, Northwestern University, Evanston, 60208, IL, USA.*
[ǂ]*Higgs Centre for Theoretical Physics, The University of Edinburgh, EH9 3FD, UK.*

*e-mail: esantos@exseed.ed.ac.uk, balicas@magnet.fsu.edu

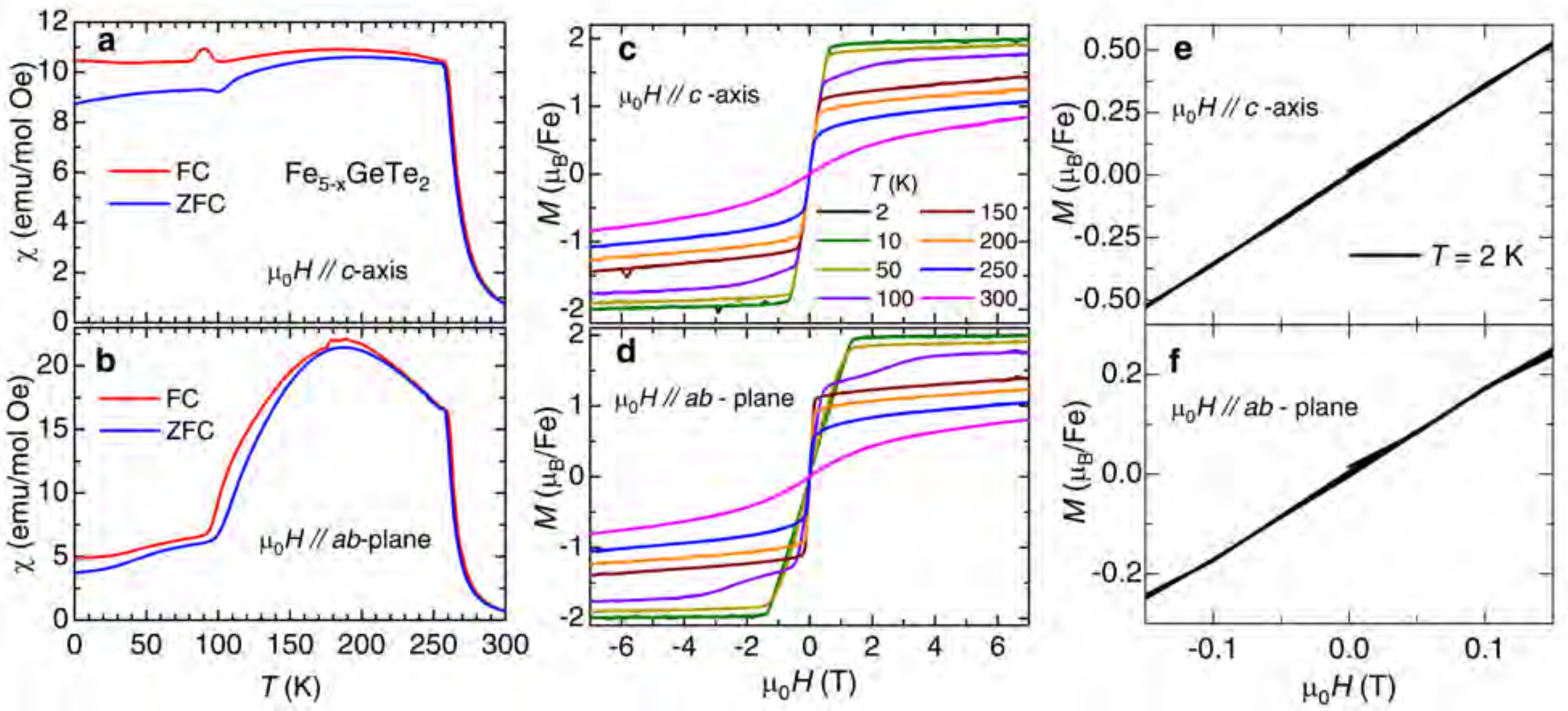

**Figure S1. Magnetic response of bulk single-crystalline $Fe_{5-x}GeTe_2$. (a) Magnetic susceptibility χ as a function of the temperature *T* measured under a field $\mu_0 H$ = 100 G applied along the *c*-axis of the crystal. (b) χ as a function of *T* for a field applied along the *ab*-plane. In both plots, red, and blue traces correspond to measurements performed under field-cooled and zero-field cooled conditions, respectively. The sharp increase in χ observed below *T* = 300 K corresponds to the Curie temperature, while an anomaly is observed in the vicinity of 100 K and due to a magneto-structural transition. (c) Magnetization *M* as a function of the $\mu_0 H$ applied along the *c*-axis, for several temperatures. (d) *M* as a function of the $\mu_0 H$ applied along the *ab*-plane, for several temperatures. (e) *M* as a function of the $\mu_0 H$ applied along the *c*-axis shown in an amplified scale around $\mu_0 H = 0$ T, at *T* = 2 K. (f) *M* as a function of the $\mu_0 H$ applied along the *ab*-plane shown in an amplified scale around $\mu_0 H = 0$ T, at *T* = 2 K. Notice the absence of hysteresis in bulk single crystals.**

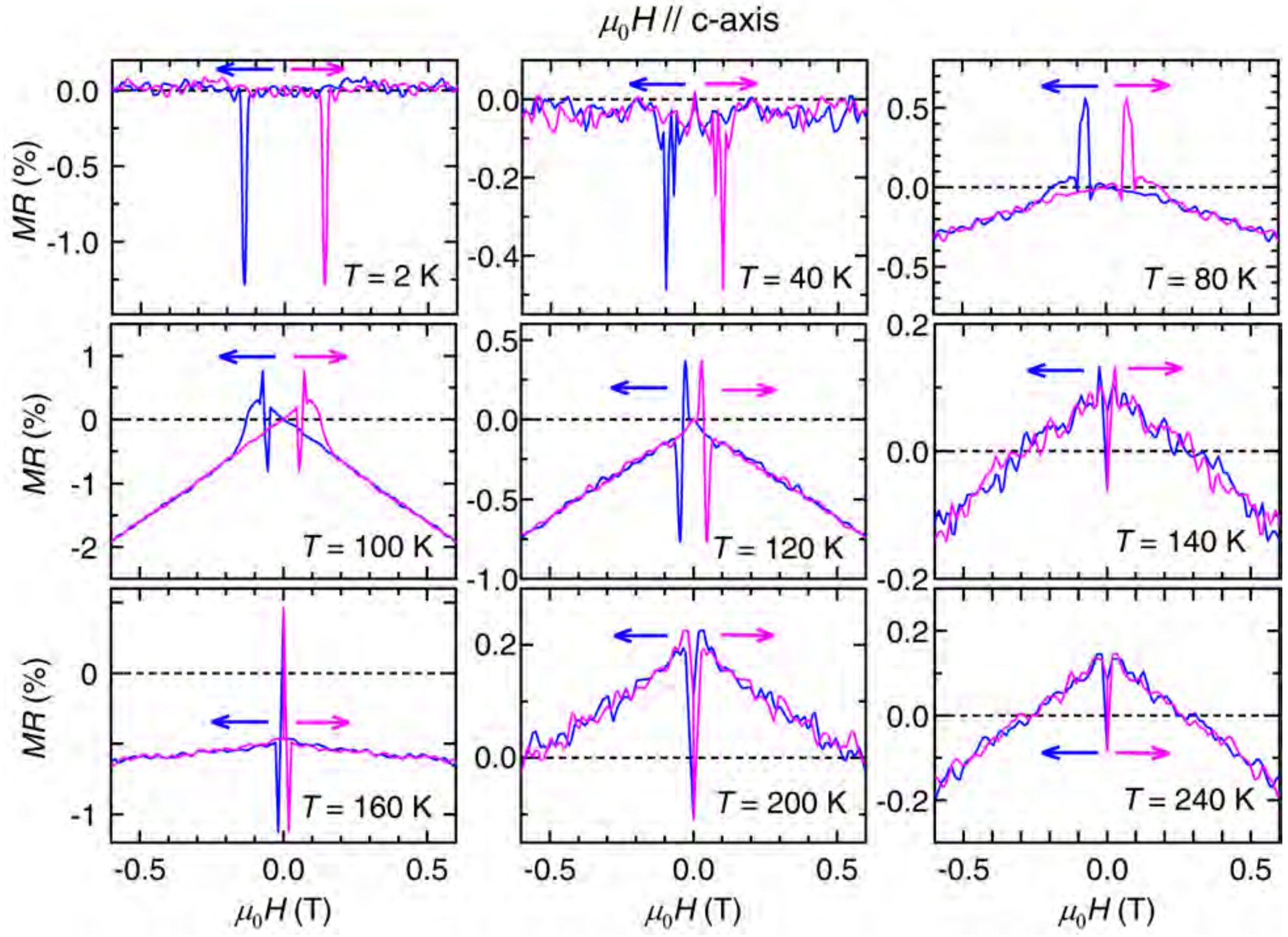

**Figure S2. Transverse magnetoresistivity $MR$ as a function of the magnetic field $\mu_0 H$ applied along the $c$-axis of a 15 nm thick $Fe_{5-x}GeTe_2$ crystal for several temperatures. Here, $MR$ = $[\rho_{xx}^c(T,\mu_0 H) - \rho_{xx}^c(T,\mu_0 H = 0\text{ T})/\rho_{xx}^c(T,\mu_0 H = 0\text{ T})]$. As indicated by horizontal arrows, magenta and blue lines correspond to increasing and decreasing magnetic field sweeps, respectively. Due to the hysteretic behavior of the $MR$ at low fields, for each trace, and after having stabilized the temperature, we took the average value of $\rho_{xx}^c(\mu_0 H = 0\text{ T})$ prior to sweeping $\mu_0 H$ along each orientation. Notice i) the very small magnetoresistivy, i.e., in the order of a few percent or a fraction of a percent, ii) the decrease in the $MR$ as a function of $\mu_0 H$, implying negative magnetoresistivity due to the field-induced suppression of spin scattering, and iii) the near absence of magnetoresistivity below $T \cong 40$ K. Below $T \cong 160$ K, sharp and hysteretic peaks, or dips, are observed in the $MR$. These coincide with the pronounced increase in $\rho_{xy}(T)$ as a function of $\mu_0 H$ prior to its saturation (see, Figures 2 and 3 in the main text). At higher $T$s both peaks converge producing dips in the vicinity of $\mu_0 H = 0$ T. Most importantly, the hysteresis associated with the position of the peaks and dips in $\rho_{xx}^c(T,\mu_0 H)$ coincide with the values of $H_c^c$ extracted from $\rho_{xy}^c(T,\mu_0 H)$ (see Figure 2 in the main text).**

**Table S1| Atomic coordinates of $Fe_{4.01}Ni_{0.68}GeTe_2$ from dingle crystal X-ray diffraction**

$Fe_{4.30}Ni_{0.39}GeTe_2$ Trigonal, $R\bar{3}m$ ($a$ = 4.037(4) Å, $c$ = 29.11(4) Å)

| | | x | y | z | $U_{iso}$*/$U_{eq}$ | Occ. (<1) |
|---|---|---|---|---|---|---|
| Fe1 | 6*c* | 0 | 0 | 0.0741(3) | 0.015(2) | 0.296(10) |
| Ni1 | 6*c* | 0 | 0 | 0.0741(3) | 0.015(2) | 0.0502(17) |
| Fe2 | 6*c* | 0 | 0 | 0.30925(9) | 0.0182(15) | 0.855 |
| Ni2 | 6*c* | 0 | 0 | 0.30925(9) | 0.0182(15) | 0.145 |
| Fe3 | 6*c* | 0 | 0 | 0.39669(11) | 0.0256(7) | 0.855 |
| Ni3 | 6*c* | 0 | 0 | 0.39669(11) | 0.0256(7) | 0.145 |
| Ge | 6*c* | 0 | 0 | 0.0086(2) | 0.019(2) | 0.5 |
| Te | 6*c* | 0 | 0 | 0.21944(5) | 0.0193(4) | |

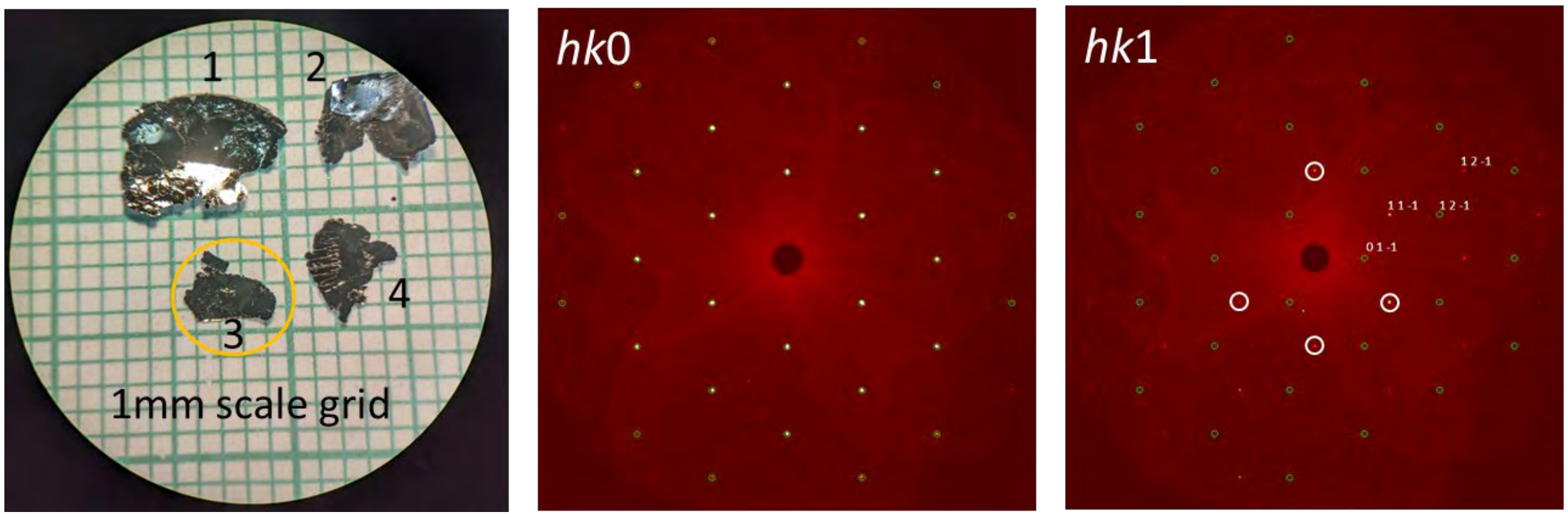

**Figure S3. Single crystal X-ray diffraction of $Fe_{4.01}Ni_{0.68}GeTe_2$. Left panel: picture of a few Ni-doped $Fe_{5-x}GeTe_2$ single crystals. Center: Precession image of the *hk*0 plane showing indexed clean bragg reflections used for refinement of the crystal structure. Right: Precession image of the *hk*1 plane indicating a set of unindexed reflections (select few in white circles) that break reflection condition - $-h + k + l = 3n$.**

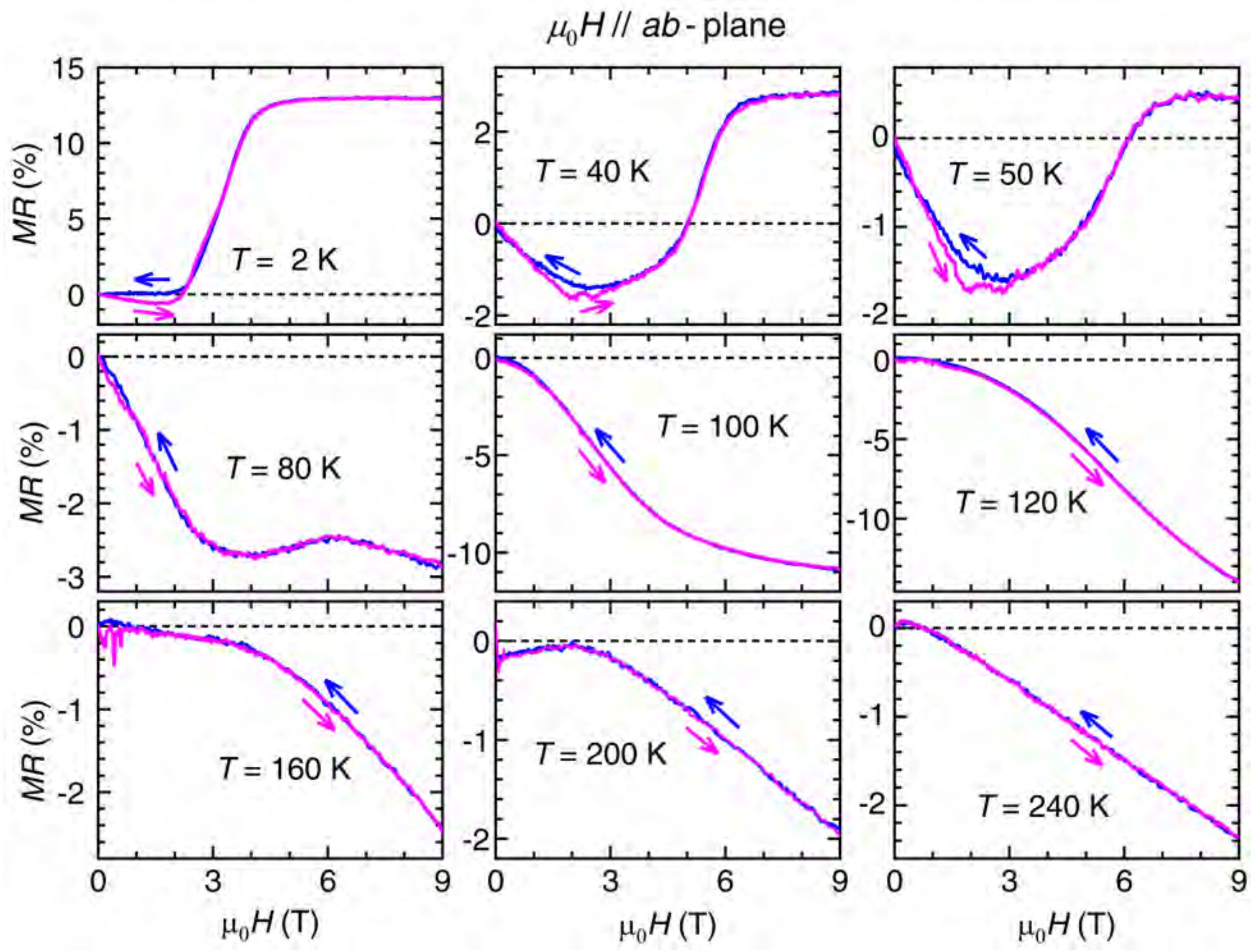

**Figure S4. Longitudinal magnetoresistivity $MR$ ($\mu_0 H \parallel j$ with $j$ being the current density) as a function of the magnetic field $\mu_0 H$ applied along the *ab*-axis for a 15 nm thick $Fe_{5-x}GeTe_2$ crystal and for several temperatures. Here, $MR = \left[\rho_{xx}^{ab}(T, \mu_0 H) - \rho_{xx}^{ab}(T, \mu_0 H = 0\ \mathrm{T}) / \rho_{xx}^{ab}(T, \mu_0 H = 0\ \mathrm{T})\right]$. As indicated by the horizontal arrows, magenta and blue lines correspond to increasing and decreasing magnetic field sweeps, respectively. Due to the hysteretic behavior of the $MR$ at low fields, for each trace, and after having stabilized the temperature, we took the average value of $\rho_{xx}^{ab}(\mu_0 H = 0\ \mathrm{T})$ prior to sweeping $\mu_0 H$ along each orientation. Notice i) how the mostly negative magnetoresistivity becomes positive beyond a certain field value as $T$ is lowered, likely due to a metamagnetic transition towards a spin-polarized state, and ii) the near absence of hysteresis in $\rho_{xx}^{ab}$ for $T \geq 100$ K, which contrasts markedly with the pronounced hysteresis observed in $\rho_{xy}^{ab}(T, \mu_0 H)$ (Figure 4 in the main text). This suggests that the hysteresis observed in $\rho_{xy}^{ab}(T, \mu_0 H)$ is not completely dominated by the motion and pinning of domain walls, which should also affect the magnetoresistivity. Instead, it suggests that the field-induced reconfiguration of spin textures is likely responsible for $\rho_{xy}^{ab}(T, \mu_0 H) \neq 0$ at $\mu_0 H = 0$ T, when $T \geq 100$ K.**

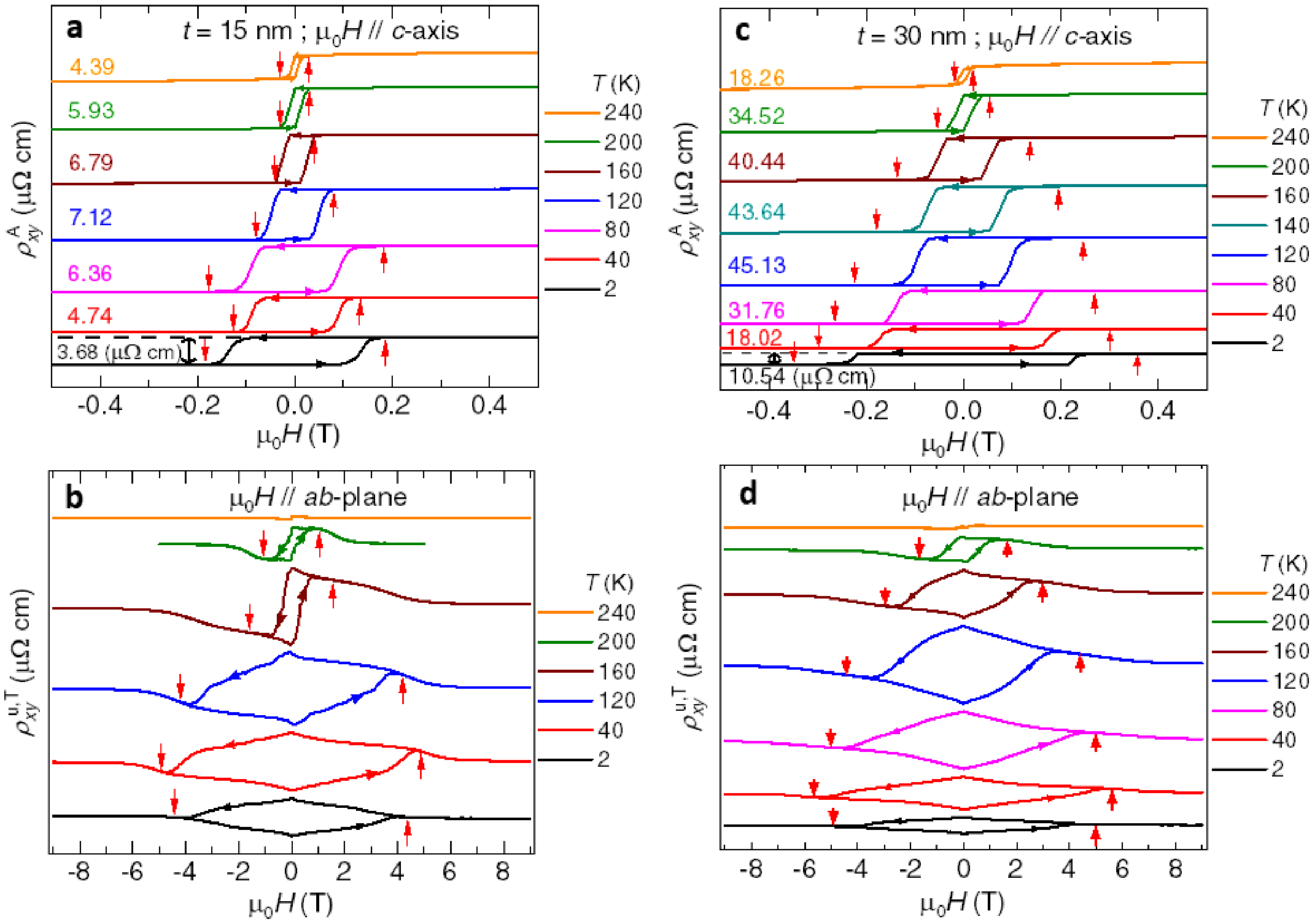

**Figure S5. Hysteretic response in both the anomalous Hall and unconventional topological Hall response of exfoliated $Fe_{5-x}GeTe_2$ crystals. (a) Anomalous Hall resistivity $\rho_{xy}^{A}$ for a $t$ = 15 nm thick exfoliated $Fe_{5-x}GeTe_2$ crystal as a function of $\mu_0 H$ and for several temperatures. Numbers indicate the absolute change $\Delta\rho_{xy}^{A}$ as the field is swept from positive to negative values. The maximum $\Delta\rho_{xy}^{A}$ occurs at temperatures just above the magneto-structural transition. Vertical red arrows indicate the coercive fields $H_c^c$ where the hysteresis loops close. (b) Unconventional topological Hall response $\rho_{xy}^{u,T}$ for the same sample, i.e., for $\mu_0 H \parallel j \parallel ab$-plane with $j$ being the current density, as a function of $\mu_0 H$ and for several temperatures. Vertical red arrows indicate the coercive fields $H_c^{ab}$ required to close the hysteresis loops. Notice how $H_c^{ab}$ exceeds $H_c^c$ by over one order of magnitude. (c) Same as in a), but for a $t$ = 30 nm thick crystal. Notice the increase in both the absolute value of both $\rho_{xy}^{A}$ and $H_c^c$. (d) Same as in (b), but for the $t$ = 30 nm thick crystal. Notice how the reversible part of $\rho_{xy}^{u,T}$ extends to higher fields. This, and similar data, was used to build Figure 5 within the main text.**

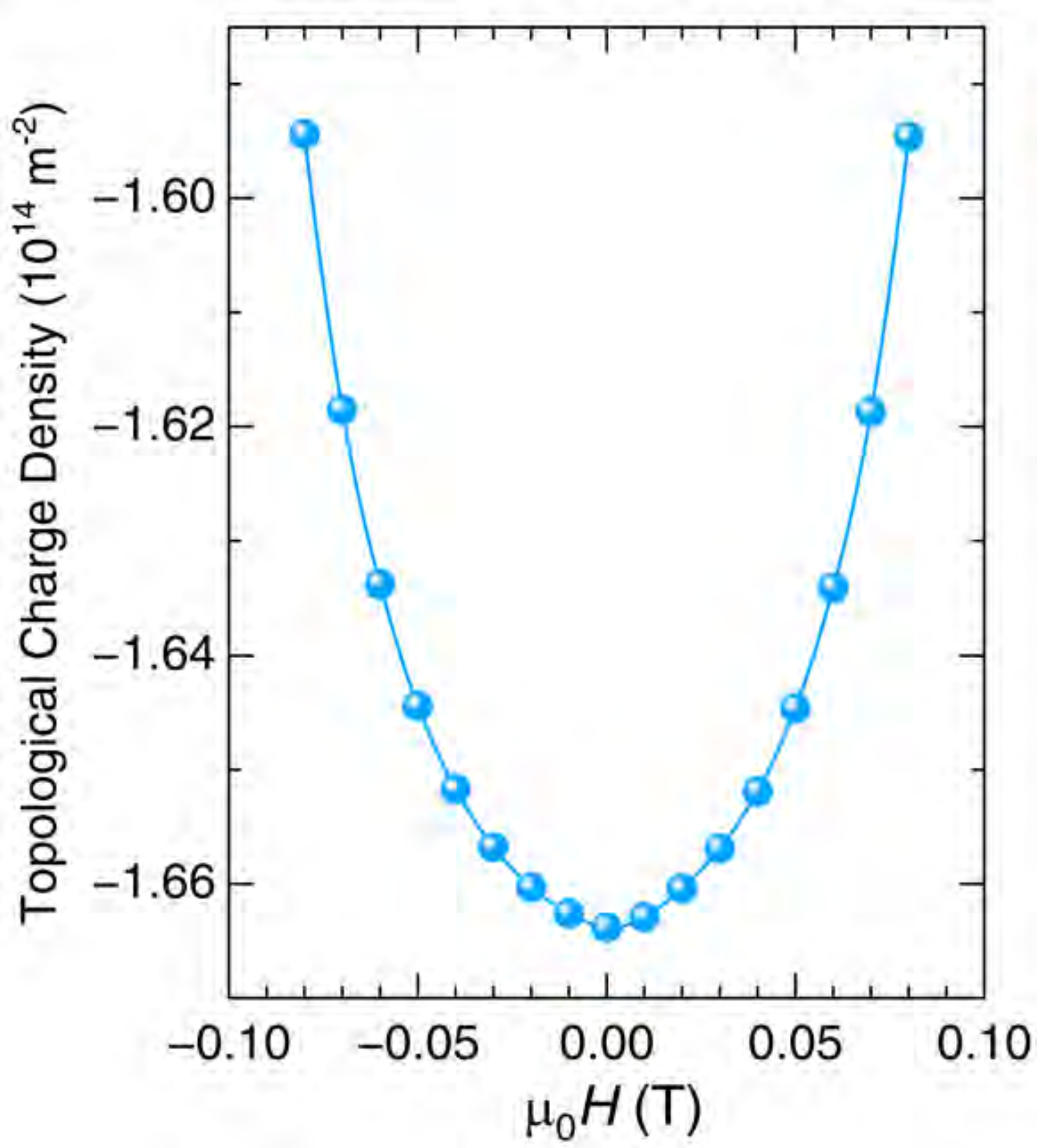

**Figure S6. Topological charge density as a function of magnetic field according to micromagnetic simulations. Notice that the absolute value of the topological charge density peaks at zero-field (maximum meron density), precisely where the unconventional topological Hall response displays its maximum value. External magnetic fields applied along an in-plane direction orients the ferromagnetic domains, thus suppressing the topological charges associated to merons.**

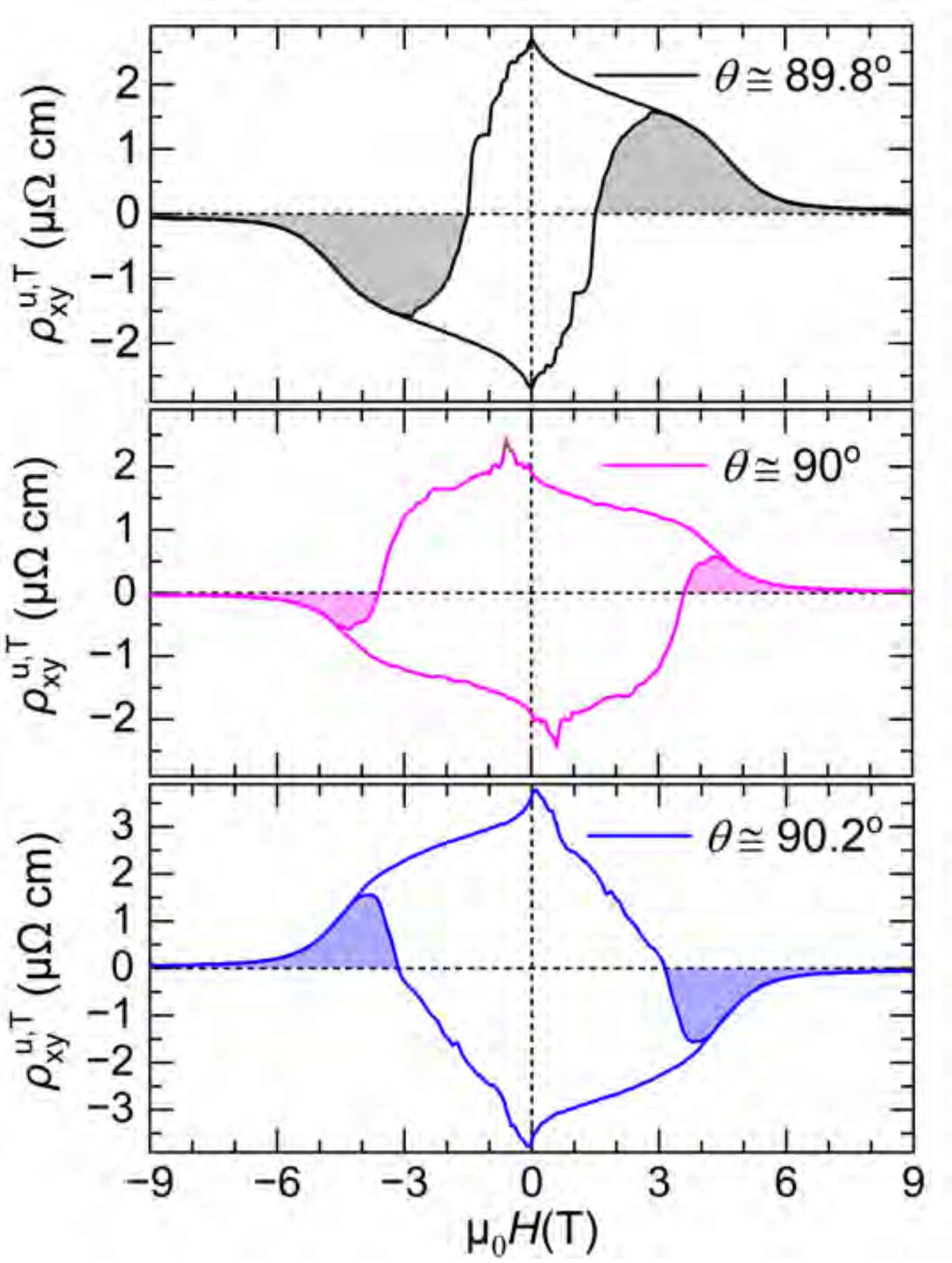

**Figure S7. Unconventional topological Hall response for fields very close to the *ab*-plane. From top to bottom we display $\rho_{xy}^{u,T}(\mu_0 H)$ for a few angles very close to the *ab*-plane at $T$ = 4 K. Notice how the pronounced broad peaks, beyond the irreversible region, change sign indicating that the topological spin textures are written by the out-of-plane component of the magnetic field. Since the $\rho_{xy}^{u,T}(\mu_0 H)$ changes sign, it should reach zero value for fields exactly along a planar direction. However, this precise alignment is very difficult to achieve with conventional mechanical rotators, hence the finite $\rho_{xy}^{u,T}(\mu_0 H)$ response at the nominal value $\theta$ = 90°.**

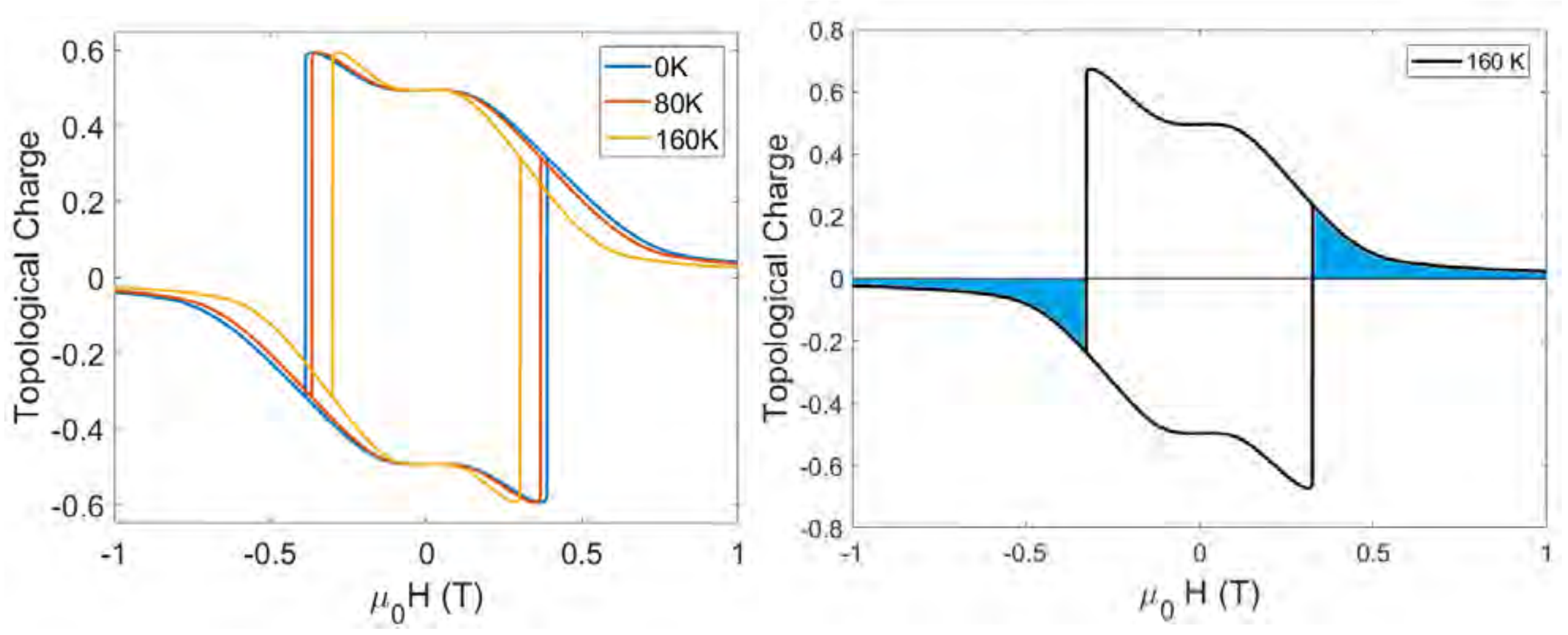

**Figure 8**: **Calculation of the topological charge as a function of inter-planar magnetic field**. **Left: Topological charge as a function of an out-of-plane field $\mu_0 H(T)$ with an absent out-of-plane anisotropy at different temperatures (0 K, 80 K, 160 K), which led to the formation of merons. The width of the loop decreases as function of the temperature as expected, depending on the reduction of the overall magnetization. Right: Similar as Right plot for 160 K with the highlighted areas of finite topological charge**.

| | $M_s$ | A | $K_1$ | D |
|---|---|---|---|---|
| 0K | $6.3 \times 10^5$ A/m | $1.0 \times 10^{-11}$ J/m | $2.5 \times 10^5$ J/m$^3$ | $1.2 \times 10^{-3}$ J/m$^2$ |
| 80K | $5.97 \times 10^5$ A/m | $8.98 \times 10^{-12}$ J/m | $2.29 \times 10^5$ J/m$^3$ | $1.07 \times 10^{-3}$ J/m$^2$ |
| 160K | $4.88 \times 10^5$ A/m | $6.02 \times 10^{-12}$ J/m | $1.85 \times 10^4$ J/m$^3$ | $7.2 \times 10^{-4}$ J/m$^2$ |

**Table S2. Parameters used for the micromagnetic simulations. Thermal parameters used in the simulations for Fig. S8 from Refs.[1-3]. $M_s$, $A$, $K_1$, and $D$ are the magnetization saturation, exchange constant, axial anisotropy, and $D$ the Dzyaloshinskii–Moriya interaction constant, respectively.**